\documentclass[11pt]{article}
\usepackage{axodraw,mathrsfs}

\catcode`\@=11
\@addtoreset{equation}{section}
\catcode`\@=12

\newcommand{\plim}{{\parbox{0pt}{\hbox{$\scriptstyle E_f' = E_f$}
\hbox{$\scriptstyle \vec P\,'\rightarrow \vec P$}}}}

\newcommand{\lplim}{{\parbox{0pt}{\hbox{$\scriptstyle E^\prime_\ell = E_\ell$}
\hbox{$\scriptstyle \vec P\,'\rightarrow \vec P$}}}}

\newcommand{\Qlim}{{\parbox{0pt}{\hbox{$\scriptstyle q^0 = 0$}
\hbox{$\scriptstyle \vec Q\rightarrow \vec 0$}}}}

\title{
\centerline{\normalsize hep-ph/0104248 \hfill SINP/TNP/01-9}\bigskip
\textbf{Gravitational couplings of charged leptons in a medium}
}
 
\author{Indrajit Mitra$^a$\footnote{indra@theory.saha.ernet.in}, 
Jos\'e F. Nieves$^b$\footnote{nieves@ltp.upr.clu.edu} and 
Palash B. Pal$^a$\footnote{pbpal@tnp.saha.ernet.in}\\
$^a$ Saha Institute of Nuclear Physics, 1/AF Bidhan-Nagar, 
Calcutta 700064, India\\ 
$^b$ Laboratory of Theoretical Physics, 
Department of Physics, P.O. Box 23343\\
University of Puerto Rico, R\'{\i}o Piedras,
Puerto Rico 00931-3343}

\date{}

\begin{document}
\maketitle

\begin{abstract}

We calculate the leading order matter-induced corrections to the
gravitational interactions of charged leptons and their antiparticles
in a medium that contains electrons but not the other charged leptons,
such as normal matter. The gravitational coupling, which is universal
at the tree level, is found to be flavor-dependent, and also different
for the corresponding antiparticles, when the corrections of
$O(\alpha)$ are taken into account.  General expressions are obtained
for the matter-induced corrections to the gravitational mass in a
generic matter background, and explicit formulas for those corrections
are given in terms of the macroscopic parameters of the 
medium for particular conditions of the background gases.

\end{abstract}
\section{Introduction}
\label{s:intro}
The gravitational interactions are universal in the sense that the
ratio of the inertial and the gravitational masses of any particle is
a constant.  This fact, expressed in the form equivalence principle,
is one of the basic axioms of the general theory of relativity.
Although this is a feature of the theory at the classical level, it
has been shown by Donoghue, Holstein and Robinett
(DHR) \cite{DHR84,DHR86}, that the corresponding linearized quantum
theory of gravity respects this ratio, at least to $O(\alpha)$.

However, in the same series of works, it was shown that this property
is lost when the particles are in the presence of a thermal background
rather than the vacuum.  To arrive at this idea, the inertial and the
gravitational masses must be defined in the context of quantum field
theory. We consider in Sec.~\ref{sec:preliminaries} their precise
definition in terms of the particle propagator and the gravitational
vertex, which we will need in the subsequent work.  For the moment,
let us denote these two quantities by $M$ and $M'$ respectively and
summarize the results of Refs.\ \cite{DHR84,DHR86}.  The authors
calculated the corrections for the electron in a background with a
temperature $T\ll m_e$ and zero chemical potential.  Thus, the
background contained only photons, but not electrons or any other
matter particles.  The dispersion relation for an electron with
momentum $\vec P$ in the rest frame of the medium is given by
\begin{eqnarray}
E_e(P) = \sqrt{P^2 + m_e^2 + {2\over 3}\alpha\pi T^2} 
\end{eqnarray}
and $M_e = E_e(0)$. Thus, to $O(\alpha)$,
\begin{eqnarray}
M_e = m_e + {\alpha\pi T^2 \over 3m_e} \,. 
\label{DHR:M}
\end{eqnarray}
In the same reference frame, the gravitational mass was calculated to
be
\begin{eqnarray}
M'_e = M_e \left( 1 - {2 \alpha \pi T^2 \over 3M_e^2} \right) \,.
\end{eqnarray}
Using Eq.\ (\ref{DHR:M}) and keeping only terms up to $O(\alpha)$,
this can be rewritten as
\begin{eqnarray}
M'_e = m_e - {\alpha\pi T^2 \over 3m_e} \,,
\label{DHR:Mg}
\end{eqnarray}
which is different from the inertial mass.

Moreover, although in those calculations only the case of the
electron was considered explicitly, the above results are equally 
applicable to other charged fermions, such as the muon. 
In particular, we note that for any such fermion $f$, the ratio
\begin{eqnarray}
{M_f' \over M_f} = 
1 - {2\alpha\pi T^2 \over 3m_f^2} + O(\alpha^2) \,,
\end{eqnarray}
depends on the mass parameter $m_f$.  Therefore, not only the inertial
and gravitational masses of a given fermion cease to be equal when the
background effects are taken into account, but in addition the ratio
of these two quantities is no longer the same for all the particles;
i.e., universality is lost as well. This happens despite the fact that
the background contains only photons and is therefore flavor neutral.
The origin of this difference is that while the background as well as
the tree-level gravitational couplings are flavor-independent, the
mass terms in the kinetic energy part of the Lagrangian are not.  We
should not be too surprised by this fact because, for example, even in
the vacuum the anomalous magnetic moment has different contributions
for the muon and the electron, though not at the lowest order.

But in a matter background with a non-zero chemical potential, such as
the Sun or a supernova, there are contributions to gravitational mass
which are proportional to the electron and nucleon densities.  These
matter contributions can dominate over the photon-background
contribution, even when $T \ll m_e$, for which the photon contribution
becomes negligible. Moreover, the matter-induced corrections to the
gravitational mass will be different for the various charged lepton
flavors, and will not be the same for the corresponding antiparticles.

Motivated by these considerations, in this work we calculate the
leading matter-induced QED corrections to the gravitational masses of
charged fermions in a medium that consists of a photon background and
a matter background of electrons and nucleons. These represent the
dominant corrections for charged leptons and antileptons. For
strongly interacting particles such as the quarks, gluon exchange
corrections are expected to be even stronger and our results will not
apply.

Our calculation is based on the one-loop corrections to the
gravitational vertex function of the charged lepton in the
medium. Working in the context of the linearized theory of gravity, we
show in detail how the gravitational mass is determined from the
gravitational vertex function, give general expressions for the
matter-induced corrections to the gravitational mass in a generic
matter background, and give explicit formulas for the corrections in
terms of the macroscopic parameters of the background medium for a few
special cases of the background gases.

The rest of the paper is organized as follows. In
Sec.~\ref{sec:preliminaries}, we discuss the general procedure for
finding the inertial and gravitational masses. In Sec.~\ref{s:SE}, we
discuss the self-energy diagrams for the charged leptons in a medium
and find the medium-induced contributions to their inertial masses. We
also calculate the wave function normalization factors which will be
needed in the calculation of the gravitational mass later. In
Sec.~\ref{s:indgrav}, we discuss the couplings in the linearized
theory of gravity and calculate the gravitational vertex of the
leptons. In Sec.~\ref{s:calc}, we use the vertex to find the
gravitational masses of charged leptons and antileptons in a
medium. The terms involving fermion distribution functions cannot be
evaluated exactly. In Sec.~\ref{s:cases}, we evaluate the corrections
in two different limits, viz., the classical and the strongly
degenerate limit for the electron gas. Sec.~\ref{s:conclu} contains
our concluding remarks.

\section{Preliminaries}
\label{sec:preliminaries}
\subsection{Inertial mass}
The dispersion relations of the modes that propagate through the medium
are determined by solving the linear part of the effective field
equation.  For fermions that propagate with momentum $p^\mu$, this
equation, in momentum space, is
\begin{eqnarray}
\label{fieldeq}
\Big(\rlap/p - m_f - \Sigma_f(p) \Big) \xi(p) = 0 \,,
\end{eqnarray}
where $\Sigma_f$ denotes
the background-dependent part of the self-energy.
The dispersion relations of the particle and the antiparticle are
given by the positive and negative energy solutions of
Eq.\ (\ref{fieldeq}), and we denote the corresponding spinors by
$U(p) = \xi(p)$ and $V(p) = \xi(-p)$, respectively.

In an isotropic medium, the most general form of $\Sigma_f$ is
\begin{eqnarray}
\label{sigmageneral}
\Sigma_f(p) = a\rlap/p + b \rlap/v + c \,,
\end{eqnarray}
where we have introduced the vector $v^\mu$ which represents the
velocity four-vector of the medium. We will perform all calculations 
in the rest frame of the medium, in which $v^\mu$ has components
\begin{eqnarray}
v^\mu = (1, \vec 0 \,) \,,
\label{v}
\end{eqnarray}
and in that frame, we define the components of $p^\mu$ by writing
\begin{eqnarray}
p^\mu = (p^0,\vec P) \,.
\end{eqnarray}
In general, $a,b,c$ are functions of the variables $p^0$ and $P$,
which we will indicate by writing them as $a(p^0,P)$, and similarly
for the other ones, when we need to show it explicitly.  Eq.\
(\ref{sigmageneral}) can contain an additional term proportional to
$\sigma^{\mu\nu}v_\mu p_\nu$ in the more general case.  However, such
a term does not appear at the level of the one-loop calculations
\cite{weldon:fermions} that we are considering in this work, and
therefore we omit it.

Requiring Eq.\ (\ref{fieldeq}) to have non-trivial
solutions yields the condition
\begin{eqnarray}
\label{poles}
D(p^0,\vec P) = 0
\end{eqnarray}
where
\begin{eqnarray}
D(p^0,\vec P) = [(1 - a)p -bv]^2 - (m_f + c)^2 \,.
\end{eqnarray}
Eq.\ (\ref{poles}) also determines the poles of the fermion propagator
\begin{eqnarray}
S'_f(p) = \frac{1}{p - m_f - \Sigma_f} \,,
\end{eqnarray}
which can be written in the form
\begin{eqnarray}
\label{fullpropsymbolic}
S'_f = \frac{N(p_0,\vec P)}{D(p_0,\vec P)} \,,
\end{eqnarray}
with
\begin{eqnarray}
\label{Np}
N(p^0,\vec P) = (1 - a)\rlap/p -b \rlap/v + (m_f + c) \,.
\end{eqnarray}

The condition given in Eq.\ (\ref{poles}) has a positive energy solution
corresponding to the particle, given by $p^0=E_f(P)$, and a negative energy
solution corresponding to the antiparticle given by $p^0= - E_{\bar
f}(P)$, where
\begin{eqnarray}
\label{disprel}
E_{f,\bar f} (P) = \sqrt{P^2 + \left(\frac{m_f + c}{1 - a}\right)^2} 
\pm \frac{b}{1 - a} \,.
\end{eqnarray}
These are implicit equations for $E_{f,\bar f}$ as a function of
$P$.  While solving for $E_f$, for example, we need to take the quantities
$a,b,c$ appearing on the right side as functions of $E_f$ and $P$.
The corresponding inertial masses are then defined as
\begin{eqnarray}
M_{f,\bar f} = E_{f,\bar f}(0) \,.
\end{eqnarray}
Since $a,b,c$ are of $O(e^2)$, we can solve
Eq.\ (\ref{disprel}) perturbatively by substituting the tree-level
value $p^0 = \pm\sqrt{P^2 + m_f^2}$ in the right-hand side.
It is useful to introduce the notation
\begin{eqnarray}
\label{calEdef}
{\cal E}_{f, \bar f}(p^0,\vec P) \equiv (a p\cdot v + b) \pm c \,,
\end{eqnarray}
which can be expressed concisely in terms of $\Sigma_f$ as
\begin{eqnarray}
\label{calEoper}
{\cal E}_{f, \bar f} = \frac{1}{4}\mbox{Tr}\; \Big[(\rlap/v \pm
1)\Sigma_f \Big]  \,.
\end{eqnarray}
To $O(e^2)$, the inertial masses are then found to be given by 
\begin{eqnarray}
\label{Moper}
M_f & = & m_f + {\cal E}_f(m_f,\vec 0) \,, \nonumber\\
M_{\bar f} & = & m_f - {\cal E}_{\bar f} (-m_f,\vec 0) \,.
\end{eqnarray}
Equation (\ref{calEoper}) is a useful formula that allows us
to extract the matter-induced corrections to the inertial
mass directly from the one-loop expression for $\Sigma_f$.
As we will see next, the wavefunction renormalization factor is
determined in terms of the same quantities ${\cal E}_f$ and ${\cal
E}_{\bar f}$.

\subsection{Wave function}
\label{s:Z}
We consider in some detail the case of the particles, and summarize
at the end the corresponding results for the antiparticles.
We adopt the normalization of the one-particle states such that their
state vectors $|f(p,s)\rangle$ satisfy
\begin{eqnarray}
\langle f(p',s')|f(p,s)\rangle = 
(2\pi)^3 \delta^{(3)}(\vec P - \vec P')\delta_{s,s'} \,.
\end{eqnarray}
The one-particle states have associated with them the wave functions
defined by the matrix element of the field operator
\begin{eqnarray}
\label{wavefunction}
\Big< 0 \Big|\psi(x) \Big| f(p,s) \Big>
= \sqrt{Z_f(p)} \; U_s(p)e^{-ip\cdot x} \,,
\end{eqnarray}
where $U_s(p)$ satisfies the Dirac equation
\begin{eqnarray}
\label{diraceq}
\Big(\rlap/p - m_f - \Sigma_f(p) \Big) U_s(p) = 0 
\end{eqnarray}
with $p^\mu = (E_f(P),\vec P)$.
In the rest frame of the medium, the explicit form of
$U_s(p)$ can be easily worked out. Adopting that frame, and
choosing the normalization such that
\begin{eqnarray}
\label{Unorm}
U_s^\dagger(p)U_s(p) = 1 \,,
\end{eqnarray}
it then follows that the $U_s$ satisfy the spinor sum relation
\begin{eqnarray}
\label{spinsum}
\sum_s U_s(p) \overline U_s(p) = \frac{N(E_f,\vec P) }
{2\left[(1 - a)E_f - b\right]}
\end{eqnarray}
where $N$ is defined in Eq.\ (\ref{Np}).
{}From Eq.\ (\ref{diraceq}) we obtain the identity
\begin{eqnarray}
\label{GordonUorig}
\overline U_s(p) \gamma_\mu U_s(p) = \left[ {(1-a)p_\mu -bv_\mu \over
m_f  +c} \right] \; \overline U_s(p) U_s(p) 
\end{eqnarray}
which, together with Eq.\ (\ref{Unorm}) imply the relations
\begin{eqnarray}
\overline U_s(p) U_s(p) = {m_f+c \over (1-a)E_f - b}
\label{UbarU}
\end{eqnarray}
and
\begin{eqnarray}
\overline U_s(p) \gamma_\mu U_s(p) = {(1-a)p_\mu -bv_\mu \over
(1-a)E_f - b} \,.
\end{eqnarray}
In particular, in the frame specified by Eq.\ (\ref{v}),
\begin{eqnarray}
\label{GordonUPzero}
\left[\overline U(p)U(p)\right]_{\vec P = 0} & = & 1 \,,
\nonumber\\[12pt]
\left[\overline U(p)\gamma_\mu U(p)\right]_{\vec P = 0} & = & v_\mu \,.
\end{eqnarray}

The normalization factor $Z_f$ that appears in Eq.\ (\ref{wavefunction})
is determined as follows. Near the pole $p^0 = E_f(P)$,
Eq.\ (\ref{fullpropsymbolic}) reduces to
\begin{eqnarray}
\label{Sp_pole}
S'_f(p) \approx
\frac{N(E_f,\vec P)}{(p_0 - E_f)
\left(\frac{\textstyle\partial D}{\textstyle\partial p_0}
\right)_{p^0 = E_f}} \,.
\end{eqnarray}
On the other hand, we can calculate
the one-particle contribution to the thermal propagator $iS_f'(x) =
\langle T\psi(x)\overline\psi(0)\rangle$ by inserting a
complete set of states, and retaining only the matrix elements between
the vacuum state and one-particle states. Using Eq.\
(\ref{wavefunction}), we obtain
\begin{eqnarray}
S'_f(p)\bigg|_{\rm 1-particle} \approx  
{Z_f(p) \sum_s U_s(p)\overline U_s(p) \over p^0 - E_f} 
\end{eqnarray}
near the same pole. The requirement that the residues of
these two expressions coincide, then yields
\begin{eqnarray}
Z_f(p) = \left\{2 \Big[(1 - a)E_f -b \Big]
\left(\frac{\textstyle\partial D}{\textstyle\partial p_0}
\right)^{-1}\right\}_{p^0 = E_f} \,,
\end{eqnarray}
where we have used Eq.\ (\ref{spinsum}).  To the lowest order in
$e^2$, and for the particular case $\vec P = 0$ in which we are
interested, the expression reduces to
\begin{eqnarray}
\label{Zf}
Z_f = 1 + \zeta_f \,, 
\end{eqnarray}
where
\begin{eqnarray}
\label{zf}
\zeta_f = \frac{\partial{\cal E}_f}{\partial p^0} 
\Bigg|_{p^\mu = (m_f,\vec 0)} \,,
\end{eqnarray}
with ${\cal E}_f$ given by Eq.\ (\ref{calEdef}) or (\ref{calEoper}).
{}From now on whenever we omit the dependence of $Z_f$ on $p$, it is to
be understood as the quantity evaluated at $\vec P = 0$.

For the case of the antiparticle, similar considerations
apply. The wavefunction for the antiparticles is defined by
\begin{eqnarray}
\label{wavefnV}
\Big< \bar f(p,s) \Big|\psi(x) \Big| 0 \Big> = \sqrt{Z_{\bar f}(p)} \;
V_s(p)e^{ip\cdot x} \,, 
\end{eqnarray}
where $V_s(p)$ satisfies the equation
\begin{eqnarray}
\label{antidiraceq}
\Big(\rlap/p + m_f + \Sigma_f(-p) \Big) V_s(p) = 0 \,,
\end{eqnarray}
with the normalization
\begin{eqnarray}
V^\dagger_s(p)V_s(p) = 1 \,,
\end{eqnarray}
and $p^\mu = (E_{\bar f}(P),\vec P)$. 
The analogy of Eq.\ (\ref{GordonUorig}) in the present case is
\begin{eqnarray}
\label{GordonV}
\overline V_s(p) \gamma_\mu V_s(p) = -\left[ {(1-a(-p))p_\mu + b(-p)
v_\mu \over m
+c(-p)} \right] \; \overline V_s(p) V_s(p) \,.
\end{eqnarray}
Writing 
\begin{eqnarray}
Z_{\bar f} = 1 + \zeta_{\bar f}\,,
\label{Zfbar}
\end{eqnarray}
the same procedure that lead to Eq.\ (\ref{zf})
leads to the formula
\begin{eqnarray}
\zeta_{\bar f} = 
\frac{\partial{\cal E}_{\bar f}}{\partial p^0} 
\Bigg|_{p^\mu = (-m_f,\vec 0)} \,.
\label{zfbar}
\end{eqnarray}
Eqs.\ (\ref{zf}) and (\ref{zfbar}) are the formulas that we will use
for the explicit calculations in Sec.~\ref{s:indgrav}.

We will denote by $u_s$ and $v_s$ the limiting value of the spinors
$U_s$ and $V_s$ when the effects of the medium are neglected.
They satisfy the free Dirac equation in the vacuum, as well as
the relations
\begin{eqnarray}
\bar u_s\gamma_\mu u_s & = & \frac{p_\mu}{m_f}\bar u_s u_s 
\label{ugu} \\
\bar u_s u_s & = & {m_f \over E_f} \,, 
\label{unorm}
\end{eqnarray}
with similar relations for $v_s$ but with the 
substitution $p_\mu\rightarrow -p_\mu$ in the above equations.

\subsection{Gravitational mass}
The gravitational mass is a measure of the strength of the coupling of
the fermion to the graviton.  It can be determined in terms of the
fermion's vertex function for the gravitational interaction, as
follows.

We denote by $\Gamma_{\lambda\rho}(p,p')$ the
one-particle irreducible vertex function, defined such
that the matrix element of the total stress-energy tensor
operator $\widehat T_{\lambda\rho}(x)$ between incoming
and outgoing fermion states is given by
\begin{eqnarray}
\label{Gammadef}
\Big< f(p',s') \Big|\widehat T_{\lambda\rho} (0) \Big| f(p,s) \Big>
= \sqrt{Z_f(p) Z_f(p')} \;
\overline U_{s'}(p')\Gamma_{\lambda\rho}(p,p') U_s(p) \,.
\end{eqnarray}
We perform all our calculations in the linearized theory of
gravity. This means that we  write
\begin{eqnarray}
\label{glinear}
g_{\lambda\rho} = \eta_{\lambda\rho} + 2\kappa h_{\lambda\rho}\,,
\end{eqnarray}
and then $h_{\lambda\rho}$ is identified with the graviton field and
treated as a weak field.  $\kappa$ is related to Newton's constant $G$
through the equation
\begin{eqnarray}\label{fcoupling}
\kappa = \sqrt{8\pi G} 
\end{eqnarray}
to ensure that the graviton field has the correctly normalized kinetic
energy term in the Lagrangian.  We write the complete vertex function
in the form
\begin{eqnarray}
\Gamma_{\lambda\rho} = V_{\lambda\rho} + \Gamma'_{\lambda\rho} \,,
\label{V+G}
\end{eqnarray}
where $\Gamma'_{\lambda\rho}$ denotes the 1-loop contribution
and $V_{\lambda\rho}$ is the tree-level vertex function given
by~\cite{np:gravnu,lingrav} 
\begin{eqnarray}
\label{Vmunu}
V_{\lambda\rho} (p,p') = \frac14 \left[
\gamma_\lambda(p + p')_\rho + 
\gamma_\rho(p + p')_\lambda \right]
- \frac12 \eta_{\lambda\rho}
\left[(\rlap/ p - m_f) + (\rlap/ p' - m_f) \right] \,.
\end{eqnarray}

We now consider the scattering of the fermion off a static
gravitational potential, which is produced by a static mass
density $\rho^{\rm ext}(\vec x)$. Defining the Fourier transform
\begin{eqnarray}
\label{phiq}
\phi^{\rm ext}(\vec x) = \int\frac{d^3q} {(2\pi)^3} 
\phi^{\rm ext} (\vec q) e^{i\vec q\cdot\vec x} \,, 
\end{eqnarray}
with a similar definition for $\rho^{\rm ext}(\vec q)$,
the corresponding metric is such that, in momentum space,
\begin{eqnarray}
\label{hphirel}
h^{\lambda\rho}(\vec q) = \frac{1}{\kappa} \phi^{\rm ext} (\vec q) 
\left(2v^\lambda v^\rho - \eta^{\lambda\rho}\right) \,,
\end{eqnarray}
where we have used the Poisson equation $-2\vec q\,^2\phi^{\rm ext} =
\kappa^2\rho^{\rm ext}$.  The formula in Eq.\ (\ref{hphirel}) is the
solution to the linearized field equation for the metric with the
static energy momentum tensor $T^{\lambda\rho} = v^\lambda v^\rho
\rho^{\rm ext}$, where $\rho^{\rm ext}$ is independent of time.  Under
the influence of such an external potential, the on-shell
$f\rightarrow f$ transition amplitude is then
\begin{eqnarray}
\label{Sff}
S_{ff} = -i \kappa (2\pi)\delta(E_f - E'_f) \sqrt{Z_f(p) Z_f(p')}
\left[\overline U_s(p') \Gamma_{\lambda\rho}(p,p') U_s(p) \right] 
h^{\lambda\rho}(\vec P - \vec P') \,,
\end{eqnarray}
Substituting Eq.\ (\ref{hphirel}) in (\ref{Sff}) yields
\begin{eqnarray}
\label{Sff2}
S_{ff} = -i(2\pi)\delta(E_f - E'_f){\cal M}(\vec P,\vec P')
\phi^{\rm ext}(\vec P - \vec P') \,,
\end{eqnarray}
where we have defined
\begin{eqnarray}
\label{calM}
{\cal M} (\vec P,\vec P') \equiv
(2v^\lambda v^\rho - \eta^{\lambda\rho}) \sqrt{Z_f(p) Z_f(p')} 
\Big[\overline U_s(p')\Gamma_{\lambda\rho}(p,p') U_s(p)\Big]_{
E_f' = E_f} \,.
\end{eqnarray}
${\cal M} (\vec P,\vec P')$ is essentially the off-diagonal element of
the Fourier transform of the mass operator, and the gravitational mass
is simply the value of this quantity when both initial and final
fermions have vanishing 3-momentum,
\begin{eqnarray}
\label{mgrav}
M'_f \equiv \lim_{\vec P\rightarrow 0} \left[ 
{\cal M}(\vec P,\vec P')\right]_{\vec P'\rightarrow \vec P} \,.
\end{eqnarray}

To justify more fully this identification, notice that the mass
density operator for the fermion, $\rho_f(t,\vec x)$, is determined by
writing an effective Lagrangian
\begin{eqnarray}
\label{effectiveL}
{\mathscr L}_{\rm eff} = -\rho_f(t,\vec x)\phi^{\rm ext}(\vec x) 
\end{eqnarray}
such that Eq.\ (\ref{Sff2}) is reproduced by taking the S-matrix element
using $\mathscr L_{\rm eff}$ as the interaction Lagrangian. This gives
the scattering amplitude
\begin{eqnarray}
\left\langle f(p',s)\left|\int d^4x \left(i\mathscr L_{\rm eff}\right) 
\right|f(p,s)\right\rangle =  
-i2\pi\delta(E'_f - E_f)\left\langle f(p',s)\left|\rho(0,\vec
0)\right|f(p,s)\right 
\rangle\phi^{\rm ext} (\vec P - \vec P') \,.
\end{eqnarray}
Comparison with Eq.\ (\ref{Sff2}) shows that $\rho_f$ is such that
\begin{eqnarray}
\label{rho0}
\left\langle f(p',s) \left|\rho_f(0,\vec 0)\right|f(p,s)\right\rangle_{
P' = P} = {\cal M}(\vec P,\vec P')
\end{eqnarray}
with ${\cal M}(\vec P,\vec P')$ given in Eq.\ (\ref{calM}).
By definition, the gravitational mass $M'_f$ is given by
\begin{eqnarray}
\label{mgravdef}
\left[\left\langle f(p',s)\left|\int d^3x\,\rho(0,\vec
x)\right|f(p,s)\right\rangle 
\right]_{\vec P\rightarrow 0} = 
(2\pi)^3 \delta^{(3)}(\vec P - \vec P\,') M'_f
\end{eqnarray}
while, on the other hand,
\begin{eqnarray}
\label{rhome}
\left\langle f(p',s)\left|\int d^3x\rho(0,\vec
x)\right|f(p,s)\right\rangle = 
(2\pi)^3\delta^{(3)}(\vec P - \vec P')
\left\langle f(p',s) \left|\rho(0,\vec 0)\right|f(p,s)\right\rangle \,.
\end{eqnarray}
Comparing Eqs.\ (\ref{mgravdef}) and (\ref{rhome}), and using
(\ref{rho0}), we arrive at the formula given in Eq.\ (\ref{mgrav}).

\subsection{Operational definition at $O(e^2)$}
Using Eqs.\ (\ref{V+G}) and (\ref{Zf}), the formula given by Eqs.\
(\ref{calM}) and (\ref{mgrav}) can be rewritten in the form
\begin{eqnarray}
\label{M'f}
M'_f &=& (2v^\lambda v^\rho - \eta^{\lambda\rho}) \nonumber\\*
&&\times 
\lim_{\vec P\rightarrow 0}\left\{
\bigg[
\overline U_s(p') \bigg\{ 
V_{\lambda\rho}(p,p') 
+ \zeta_f V_{\lambda\rho}(p,p') 
+ Z_f \Gamma'_{\lambda\rho}(p,p') 
\bigg\} U_s(p) 
\bigg]_\plim\right\} \,.
\end{eqnarray}
Since $\zeta_f$ and $\Gamma'_{\lambda\rho}$ are $O(e^2)$, in any term
that contains either of these factors we substitute the tree level
expressions for the other quantities.  Furthermore, the terms
involving $V_{\lambda\rho}$ can be evaluated immediately with the help
of the identities given in Eq.\ (\ref{GordonUPzero}).  Remembering
that $E_f(0)=M_f$, we finally obtain the operational definition to
$O(e^2)$
\begin{eqnarray}
\label{operationalM'}
M'_f = 3M_f - 2m_f + \zeta_f m_f 
+ (2v^\lambda v^\rho - \eta^{\lambda\rho})
\lim_{\vec P\rightarrow 0}\left\{\left[
\vphantom{\frac{1}{2}}
\overline u_s(p')\Gamma'_{\lambda\rho}(p,p')u_s(p)
\right]_\plim\right\} \,,
\end{eqnarray}
where we can set $E_f = \sqrt{P^2 + m_f^2}$ in the last term.

The arguments for the case of the antiparticle 
are similar, but the equation corresponding to Eq.\ (\ref{Sff}) 
is
\begin{eqnarray}
\label{Sbarff}
S_{\bar f\bar f} = (-1)(-i \kappa)(2\pi)
\delta(E_{\bar f} - E'_{\bar f}) \sqrt{Z_{\bar f}(p) Z_{\bar f}(p')} \;
\left[\overline V_s(p) \Gamma_{\lambda\rho}(-p',-p) V_s(p') 
\right] h^{\lambda\rho}(\vec P - \vec P') \,,
\end{eqnarray}
where the extra minus sign is due to the usual fermion exchange rule.
This leads to an equation that is analogous to Eq.\ (\ref{M'f}),
but with an extra minus sign in front and some obvious changes
in the corresponding symbols, which in turn lead to the $O(e^2)$ formula
\begin{eqnarray}
\label{opMbar'}
M'_{\bar f} & = & 3M_{\bar f} - 2m_f + \zeta_{\bar f} m_f 
- (2v^\lambda v^\rho - \eta^{\lambda\rho})
\lim_{\vec P\rightarrow 0}\left\{\left[
\vphantom{\frac{1}{2}}
\overline v_s(p)\Gamma'_{\lambda\rho}(-p',-p)v_s(p')
\right]_\plim\right\} .\;
\end{eqnarray}
For the following discussion, it is useful to indicate explicitly the
dependence of the vertex function $\Gamma'_{\lambda\rho}(p,p')$ on the
vector $v^\mu$, and therefore we will write as
$\Gamma'_{\lambda\rho}(p,p',v)$.  Using the usual relation between the
free particle and antiparticle spinors by means of the charge
conjugation matrix $C$, the spinor matrix element that appears in Eq.\
(\ref{opMbar'}) can be rewritten in the form
\begin{eqnarray}
\label{Crelation}
\overline v_s(p)\Gamma'_{\lambda\rho}(-p',-p,v)v_s(p') = 
-\overline u_s(p')\Gamma'^{\,c}_{\lambda\rho}(-p',-p,v)u_s(p) \,,
\end{eqnarray}
where, for any $4\times 4$ matrix $A$, we define
\begin{eqnarray}
A^c \equiv C^{-1}A^T C \,.
\end{eqnarray}
On the other hand, the following result holds. If the Lagrangian of
the theory is $C$ invariant (which in our case it is) and if
the background is $C$-symmetric, then the gravitational vertex
function satisfies the relation
\begin{eqnarray}
\label{Cinv}
\Gamma'^{\,c}_{\lambda\rho}(-p',-p,v) =
\Gamma'_{\lambda\rho}(p,p',v) \,.
\end{eqnarray}
This result is obtained by the same techniques that were employed in
Ref.\ \cite{np:nueprop} to analyze the transformation properties of
the induced electromagnetic vertex of neutrinos in a matter
background. This result cannot be applied in our case because we will
consider backgrounds which are not particle-antiparticle
asymmetric. However, as an extension of Eq.\ (\ref{Cinv}), similar
arguments can be used to show that, if the Lagrangian is $C$ invariant
but the background is not $C$-symmetric, then the vertex function
satisfies
\begin{eqnarray}
\label{nonCback}
\Gamma'^{\,c}_{\lambda\rho}(-p',-p,v) =
\Gamma'_{\lambda\rho}(p,p',-v) \,.
\end{eqnarray}
Using Eq.\ (\ref{nonCback}) in (\ref{Crelation}) and substituting
the result in Eq.\ (\ref{opMbar'}), we then obtain
the formula
\begin{eqnarray}
\label{CoperationalMbar'}
M'_{\bar f} = 3M_{\bar f} - 2m_f + \zeta_{\bar f} m_f 
+ (2v^\lambda v^\rho - \eta^{\lambda\rho})
\lim_{\vec P\rightarrow 0}\left\{\left[
\vphantom{\frac{1}{2}}
\overline u_s(p')\Gamma'_{\lambda\rho}(p,p',-v)u_s(p)
\right]_\plim\right\} \,.
\end{eqnarray}

We take the opportunity to emphasize the following point. In the
calculations that follow, we will find expressions for the various
contributions to $\Gamma'_{\lambda\rho}(p,p')$, which are given as
integrals over the propagators and thermal distribution functions. In
general, such expressions do not have a unique limiting value as we
let $p'\rightarrow p$ in an arbitrary way
\cite{zeromomprob}. Moreover, some of the integrals are ill-defined if
the limit is not taken properly.  In our case, the precise order in
which the various limits must be taken has been dictated by the
physical issue at hand. Thus, since we are interested in the
interaction of the particle with a static gravitational potential, the
quantity that enters is $\Gamma'_{\lambda\rho}(p,p')$, evaluated for
$E'_f=E_f$.  Next we set $\vec P' = \vec P$ since we
actually want the forward scattering amplitude, and finally set $\vec
P\rightarrow 0$ to obtain the coupling at zero momentum, which
determines the gravitational mass. This justifies the somewhat
cumbersome notation regarding the limits in Eq.\
(\ref{operationalM'}), but it is meant to indicate precisely what we
have just explained, since failure to follow this prescription results
in ill-defined expressions in some contributions. On the other hand,
as we will see, this prescription allowed us to evaluate all the
integrals involved, in a unique and well-defined way, including those
that superficially seem to be singular, without having to introduce by
hand any special regularization technique.

\section{Self-energy}
\label{s:SE}
\subsection{Calculation of ${\cal E}_\ell$}
\begin{figure}
\begin{center}
%
%
\begin{picture}(180,130)(-90,-30)
\Text(0,-30)[c]{\large\bf (A)}
\ArrowLine(80,0)(40,0)
\Text(60,-10)[c]{$\ell(p)$}
\ArrowLine(40,0)(-40,0)
\Text(0,-10)[c]{$\ell(p+k)$}
\ArrowLine(-40,0)(-80,0)
\Text(-60,-10)[cr]{$\ell(p)$}
\PhotonArc(0,0)(40,0,180){4}{6.5}
\Text(0,50)[cb]{$\gamma(k)$}
\end{picture}
%
%
\begin{picture}(100,100)(-50,-30)
\Text(0,-30)[c]{\large\bf (B)}
\ArrowLine(40,0)(0,0)
\Text(35,-10)[cr]{$\ell(p)$}
\ArrowLine(0,0)(-40,0)
\Text(-35,-10)[cl]{$\ell(p)$}
\Photon(0,0)(0,35){2}{6}
\Text(-4,20)[r]{$\gamma$}
\ArrowArc(0,55)(20,-90,270)
\Text(0,85)[b]{$f(k)$}
\end{picture}
\caption[]{\sf One-loop diagrams for the self-energy of a charged
lepton $\ell$ in a medium. 
\label{f:selfenergy}}
\end{center}
\end{figure}
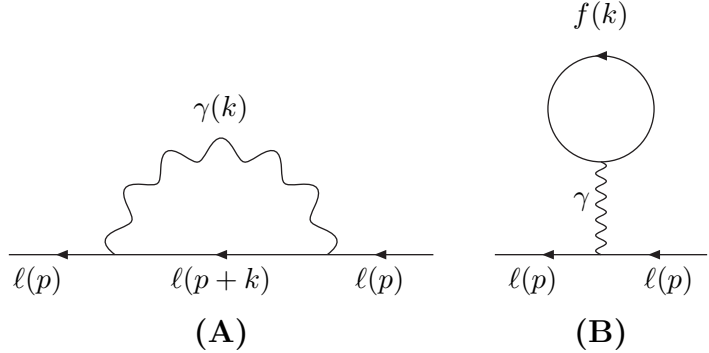

The self-energy diagrams are shown in Fig.~\ref{f:selfenergy}.  In the
absence of a gravitational potential, the contribution from
Fig.~\ref{f:selfenergy}B vanishes because the photon tadpole is zero
in an electrically neutral medium~\cite{NP92}.  In the presence of a
gravitational potential, that diagram is not zero by itself because
the condition for the vanishing of the photon tadpole, which is
equivalent to require that the medium be electrically neutral,
involves other diagrams. This will be discussed in detail in
Sec.~\ref{s:Ztype}.  As shown there, the conclusion remains that we
need to consider only Fig.~\ref{f:selfenergy}A to calculate the
self-energy.

Therefore, the charged lepton self-energy is given by
\begin{eqnarray}
-i\Sigma_\ell(p) = (ie)^2 \int {d^4k\over (2\pi)^4} \; \gamma^\mu
iS_{\ell}(p+k)\gamma^\nu iD_{\mu\nu}(k)  \,,
\label{Sigmap}
\end{eqnarray}
where $S_\ell(k)$ and $D_{\mu\nu}(k)$ are the thermal propagators for
the internal lines.  For a fermion, the propagator is given by
\begin{eqnarray}\label{S}
iS_f(p) = iS_{Ff}(p) + S_{Tf}(p) 
\end{eqnarray}
where
\begin{eqnarray}\label{S0}
S_{Ff} &=& \frac{\rlap/{p} + m_f}{p^2 - m_f^2 + i\epsilon} \,, \\
\label{S'}
S_{Tf}(p) &=& - 2\pi (\rlap/{p} + m_f) \delta(p^2 - m_f^2) \eta_f(p) \,.
\end{eqnarray}
For  the photon, in the Feynman gauge,
\begin{eqnarray}\label{D}
iD_{\lambda\rho}(k) = -\eta_{\lambda\rho} \left[ i\Delta_F(k) +
\Delta_T(k) \right] \,, 
\end{eqnarray}
where
\begin{eqnarray}\label{D0}
\Delta_F(k) &=& {1 \over k^2+i\epsilon} \,, \\ 
\label{D'}
\Delta_T(k) &=& 2\pi \delta(k^2) \eta_\gamma(k) \,.
\end{eqnarray}
We have introduced the notation
\begin{eqnarray}\label{etaf}
\eta_f(p) &=& \frac{\theta(p\cdot v)}{e^{\beta(p\cdot v - \mu_f)} + 1}
+ \frac{\theta(-p\cdot v)}{e^{-\beta(p\cdot v - \mu_f)} + 1} \,,
\\ 
%
\label{etab}
\eta_\gamma(k) &=& \frac{1}{e^{\beta\, |k\cdot v|} - 1} \,.
\end{eqnarray}
where $\beta = 1/T$ is the inverse temperature of the background and
$\mu_f$ the chemical potential.

When Eqs.\ (\ref{S}) and (\ref{D}) are substituted into Eq.\
(\ref{Sigmap}), four terms are produced.  Since we are interested in
the background induced contributions only, we disregard the term
involving both $S_{F\ell}$ and $\Delta_F$. Among the other three, the
one involving both $S_{T\ell}$ and $\Delta_T$ contributes only to the
absorptive part of the self-energy --- i.e., to the imaginary part of
the coefficients $a,b,c$ in Eq. ({\ref{sigmageneral}) --- and
therefore do not contribute to the mass. The contributions to the real
part of the coefficients arises from the remaining two terms, which
can be written in the form
\begin{eqnarray}
\label{Sigma1,2}
\Sigma'_\ell(p) = \Sigma'_{\ell1}(p) + \Sigma'_{\ell2}(p) \,,
\end{eqnarray}
where
\begin{eqnarray}
\label{Sigma1}
\Sigma'_{\ell1}(p) & = & 
2e^2 \int {d^4k \over (2\pi)^3}\;\delta(k^2) \eta_\gamma(k)
{\rlap/p + \rlap/ k - 2m_\ell \over p^2 + 2k\cdot p - m^2_\ell}\nonumber\\
\label{Sigma2}
\Sigma'_{\ell2}(p) & = & 
-\, 2e^2 \int {d^4k \over (2\pi)^3} \; 
\delta(k^2 - m_\ell^2) \eta_\ell(k) \;
{\rlap/ k - 2m_\ell \over p^2 - 2k\cdot p + m_\ell^2} \,.
\end{eqnarray}
Using Eq.\ (\ref{calEoper}), and according to the decomposition given
in Eq.\ (\ref{Sigma1,2}), we write
\begin{eqnarray}
\label{Edecomp}
{\cal E}_\ell = {\cal E}_{\ell1} + {\cal E}_{\ell2}\,,
\end{eqnarray}
where
\begin{eqnarray}
\label{E1}
{\cal E}_{\ell1} & = & 2e^2 \int\frac{d^4k}{(2\pi)^3}
\delta(k^2) \eta_\gamma(k)\frac{p\cdot v + k\cdot v - 2m_\ell}
{p^2 + 2k\cdot p - m_\ell^2} \,,\\
\label{E2}
{\cal E}_{\ell2} & = & -2e^2 \int\frac{d^4k}{(2\pi)^3} 
\delta(k^2 - m_\ell^2) \eta_\ell(k)\frac{k\cdot v - 2m_\ell}
{p^2 - 2k\cdot p + m_\ell^2} \,.
\end{eqnarray}
We can make a similar decomposition of ${\cal E}_{\bar\ell}$. The
quantities ${\cal E}_{\bar\ell1}$ and ${\cal E}_{\bar\ell2}$ are
obtained from ${\cal E}_{\ell1}$ and ${\cal E}_{\ell2}$ by replacing
$m_\ell$ by $-m_\ell$.

\subsection{Inertial mass}
The inertial mass is determined by applying Eq.\ (\ref{Moper}) and,
according to the decomposition given in Eq.\ (\ref{Edecomp}) we
write it as
\begin{eqnarray}
\label{Melldef}
M_\ell = m_\ell + m_{\ell 1} + m_{\ell 2} \,,
\end{eqnarray}
and similarly for the anti-leptons.  Substituting $p^\mu =
(m_\ell,\vec 0)$ in Eq.\ (\ref{E1}), and using the fact that the terms
in the integrand that are odd in $k$ yield zero, we obtain
\begin{eqnarray}
m_{\ell 1} \equiv {\cal E}_{\ell 1}(m_\ell,\vec 0) & =  &
\frac{e^2}{m_\ell}\int\frac{d^4k}{(2\pi)^3}\delta(k^2)\eta_\gamma(k)
\nonumber\\
\label{m1}
& = & \frac{e^2T^2}{12m_\ell} \,.
\end{eqnarray}
This is the contribution to the inertial mass from the photons in the
background, in agreement with the result quoted in Eq.\ (\ref{DHR:M}),
and it is non-zero for any the charged lepton propagating through the
medium.  In a similar fashion we find
\begin{eqnarray}
m_{\bar\ell 1} \equiv -{\cal E}_{\bar\ell 1}(-m_\ell,\vec 0) = 
\frac{e^2T^2}{12m_\ell} \,,
\end{eqnarray}
and therefore the photon contribution for the anti-particle is the same
as for the corresponding particles.

The term given in Eq.\ (\ref{E2}) is due to the fermions in the
background. Therefore in a background that contains electrons but not
the other charged leptons, the distribution functions for the muon and
the tau vanish. As a result,
\begin{eqnarray}
m_{\mu 2} = m_{\tau 2} = m_{\bar\mu 2} = m_{\bar\tau 2} = 0 \,.
\end{eqnarray}
For the electron, we obtain
\begin{eqnarray}
m_{e2}\equiv {\cal E}_{e2}(m_e,\vec 0) =
\frac{e^2}{m_e}\int\frac{d^4k}{(2\pi)^3}\delta(k^2 - m_e^2) 
\eta_e(k)\left[\frac{k_0 - 2m_e}{k_0 - m_e}\right]
\end{eqnarray}
Performing the integrations over $k_0$ and the angular variables,
we obtain
\begin{eqnarray}
\label{me2}
\label{E+e2}
m_{e2} = 
\frac{e^2}{2\pi^2 m_e} \int_0^\infty dK \frac{K^2}{2E_K}
\left[\left({E_K - 2m_e \over
E_K - m_e} \right) f_e(E_K) + \left( {E_K + 2m_e \over
E_K + m_e} \right) f_{\bar e} (E_K) \right] \,,
\end{eqnarray}
where we have put
\begin{eqnarray}
\label{EK}
k^\mu = (E_K,\vec K) \,, \qquad E_K \equiv \sqrt{K^2 + m_e^2} \,,
\end{eqnarray}
and the distribution functions for a fermion and antifermion
are given by the usual formulas
\begin{eqnarray}
\label{distfunctions}
f_{f,\bar f}(E) = \frac{1}{e^{\beta(E \mp \mu_f)} + 1} \,,
\end{eqnarray}
respectively. Similarly,
\begin{eqnarray}
\label{mbare2}
m_{\bar e 2} & \equiv & -{\cal E}_{\bar e2}(-m_e,\vec 0) \nonumber\\
& = &
\frac{e^2}{2\pi^2 m_e} \int_0^\infty dK \frac{K^2}{2E_K}
\left[\left({E_K + 2m_e \over
E_K + m_e} \right) f_e(E_K) + \left( {E_K - 2m_e \over
E_K - m_e} \right) f_{\bar e} (E_K) \right] \,.
\end{eqnarray}

The integration over $K$ can be performed only when the momentum
distribution functions are specified, and we will consider some
examples in Sec.~\ref{s:cases}. Here we only note that, as it is
expected on the basis of $CPT$-symmetry considerations, the inertial
mass correction is the same for particle and anti-particle if the
medium has zero chemical potential, but not otherwise.

\subsection{Calculation of $Z_\ell$}
We decompose
\begin{eqnarray} 
\zeta_\ell = \zeta_{\ell 1} + \zeta_{\ell 2}
\end{eqnarray}
with a similar decomposition for the anti-leptons, where
\begin{eqnarray}
\zeta_{\ell i} = 
\frac{\partial{\cal E}_{\ell i}}{\partial p^0} \Bigg|_{
p^\mu = (m_\ell,\vec 0)} \qquad \mbox{for $i = 1,2$},
\end{eqnarray}
and
\begin{eqnarray}
\zeta_{\bar\ell i} = 
\frac{\partial{\cal E}_{\bar\ell i}}{\partial p^0}\Bigg|_{
p^\mu = (-m_\ell,\vec 0)} \qquad \mbox{for $i = 1,2$}.
\end{eqnarray}
Taking the derivative in Eq.\ (\ref{E1})
and then setting $p^\mu = (\pm m_\ell,\vec 0)$, we obtain
\begin{eqnarray}
\frac{\partial{\cal E}_{\ell 1}}{\partial p^0} \Bigg|_{
p^\mu = (m_\ell,\vec 0)} 
= 
\frac{\partial{\cal E}_{\bar\ell 1}}{\partial p^0}\Bigg|_{
p^\mu = (-m_\ell,\vec 0)} 
= -\frac{e^2}{m_\ell^2} \int\frac{d^3K}{(2\pi)^3} 
\frac{f_\gamma(K)}{K}
\left( 1 - \frac{m^2_\ell}{K^2}\right) \,,
\end{eqnarray}
which implies
\begin{eqnarray}
\label{Zpsi1}
\zeta_{\ell 1}  = \zeta_{\bar\ell 1} = 
-\, {e^2T^2\over 12m_\ell^2} + {e^2 \over 2\pi^2} \int_0^\infty
{dK\over K} \; f_\gamma(K) \,,
\end{eqnarray}
where we have introduced the photon momentum distribution function
\begin{eqnarray}
f_\gamma(K) = \frac{1}{e^{\beta K} - 1} \,.
\end{eqnarray}
The integral in Eq. (\ref{Zpsi1}) is infrared divergent,
and it will cancel a similarly divergent term in the gravitational vertex
contribution to the gravitational mass [See Eq. (\ref{A1final})].

Since the electron background terms do not contribute to the
self-energy of the muon or the tau, it follows that
\begin{eqnarray}
\zeta_{\mu 2} = \zeta_{\tau2} 
= \zeta_{\bar\mu 2} = \zeta_{\bar\tau 2} = 0 \,.
\end{eqnarray}
For the electron, Eq.\ (\ref{E2}) implies
\begin{eqnarray}
\frac{\partial{\cal E}_{e2}}{\partial p^0}\Bigg|_{
p^\mu = (m_e,\vec 0)} & = & - {{\cal E}_e (m_e,\vec 0) \over m_e}
\nonumber\\ 
\frac{\partial{\cal E}_{\bar e2}}{\partial p^0}\Bigg|_{
p^\mu = (-m_e,\vec 0)} & = & {{\cal E}_{\bar e}(-m_e,\vec 0) \over
m_e} \,, 
\end{eqnarray}
which yield
\begin{eqnarray}
\label{ZU2}
\label{Le2}
\zeta_{e2} & = & - \frac{m_{e2}}{m_e}\nonumber\\
\zeta_{\bar e 2} & = & - \frac{m_{\bar e 2}}{m_e} \,,
\end{eqnarray}
with $m_{e2}$ and $m_{\bar e 2}$ given in Eqs.\ (\ref{me2}) and 
(\ref{mbare2}), respectively.

\section{Gravitational vertex}
\label{s:indgrav}
\subsection{Irreducible diagrams and couplings}
The irreducible one-loop diagrams for the vertex function are given in
Figs.~\ref{f:Wtype} and \ref{f:Ztype}.  We adopt the convention that
$q$ is the momentum of the outgoing graviton, so that
\begin{eqnarray}
q = p - p' \,,
\end{eqnarray}
and we calculate only the terms that contribute to the the
dispersive part of the vertex function, which satisfies the condition
\begin{eqnarray}
\Gamma_{\lambda\rho} (p,p') = \gamma_0 \Gamma_{\lambda\rho}^\dagger
(p',p) \gamma_0 \,.
\end{eqnarray}
The absorptive part contributes to the fermion damping, with which
we are not concerned in the present work.

When the formulas given in Eqs.\ (\ref{S}) and (\ref{D}) for the
propagators are substituted in the expressions corresponding to the
diagrams, we obtain terms of different kind.  One of them is
independent of the background medium, in which we are not interested.
Those involving two factors of the thermal part of the propagators
contribute to the absorptive part of the vertex, while those involving
three factors of the thermal part vanish because of the various
$\delta$-functions appearing in it. Thus, the background induced
contribution to the dispersive part of the vertex, to be denoted by
$\Gamma'_{\lambda\rho}$, contains the thermal part of only one of the
propagators, and they are the only kind of term that we retain.

We have omitted the one-particle reducible diagrams in which the
graviton line comes out from one of the external fermion legs, because
they do not contribute to $\Gamma_{\lambda\rho}$. The proper way to
take them into account in the calculation of the amplitude for any
given process, is by choosing the external spinor to be
the solution of the effective Dirac equation for the propagating
fermion mode in the medium, instead of the spinor representing the
free-particle solution of the equation in the vacuum, with the
normalization determined by the self-energy of the fermion, as
discussed in Sec.~\ref{s:Z}.
%
%
\begin{figure}[tbp]
\begin{center}
%
%
\begin{picture}(180,150)(-90,-65)
\Text(35,-35)[ct]{\large\bf (A)}
\ArrowLine(80,0)(40,0)
\Text(60,-10)[c]{$\ell(p)$}
\ArrowLine(40,0)(0,0)
\Text(20,-10)[c]{$\ell(k)$}
\ArrowLine(0,0)(-40,0)
\Text(-20,-10)[c]{$\ell(k-q)$}
\ArrowLine(-40,0)(-80,0)
\Text(-60,-10)[c]{$\ell(p')$}
\Photon(0,0)(0,-45){2}{4}
\Photon(0,0)(0,-45){-2}{4}
\Text(0,-50)[l]{$q$}
\PhotonArc(0,0)(40,0,180){4}{7.5}
\Text(0,50)[c]{$\gamma$}
\end{picture}
%
%
\begin{picture}(180,130)(-90,-65)
\Text(0,-35)[ct]{\large\bf (B)}
\ArrowLine(80,0)(40,0)
\Text(60,-10)[c]{$\ell(p)$}
\ArrowLine(40,0)(-40,0)
\Text(0,-10)[c]{$\ell(p-k)$}
\ArrowLine(-40,0)(-80,0)
\Text(-60,-10)[cr]{$\ell(p')$}
\Photon(0,44)(0,80){2}{4}
\Photon(0,44)(0,80){-2}{4}
\Text(0,85)[bl]{$q$}
\PhotonArc(0,0)(40,0,180){4}{6.5}
\Text(40,40)[c]{$\gamma$}
\Text(-40,40)[c]{$\gamma$}
\end{picture}
%
%
\begin{picture}(180,130)(-90,-65)
\Text(0,-35)[ct]{\large\bf (C)}
\ArrowLine(80,0)(40,0)
\Text(60,-10)[c]{$\ell(p)$}
\ArrowLine(40,0)(-40,0)
\Text(0,-10)[c]{$\ell(k)$}
\ArrowLine(-40,0)(-80,0)
\Text(-60,-10)[c]{$\ell(p')$}
\PhotonArc(0,0)(40,0,180){4}{7.5}
\Text(0,50)[c]{$\gamma$}
\Photon(40,0)(40,-50){2}{4}
\Photon(40,0)(40,-50){-2}{4}
\Text(40,-55)[l]{$q$}
\end{picture}
%
%
%
\begin{picture}(180,130)(-90,-65)
\Text(0,-35)[ct]{\large\bf (D)}
\ArrowLine(80,0)(40,0)
\Text(60,-10)[c]{$\ell(p)$}
\ArrowLine(40,0)(-40,0)
\Text(0,-10)[c]{$\ell(k)$}
\ArrowLine(-40,0)(-80,0)
\Text(-60,-10)[c]{$\ell(p')$}
\PhotonArc(0,0)(40,0,180){4}{7.5}
\Text(0,50)[c]{$\gamma$}
\Photon(-40,0)(-40,-50){2}{4}
\Photon(-40,0)(-40,-50){-2}{4}
\Text(-40,-55)[r]{$q$}
\end{picture}
\caption[]{\sf 
One-loop diagrams for the gravitational vertex of charged
leptons in a background of electrons. The braided line represents the
graviton.
\label{f:Wtype}}
\end{center}
\end{figure}
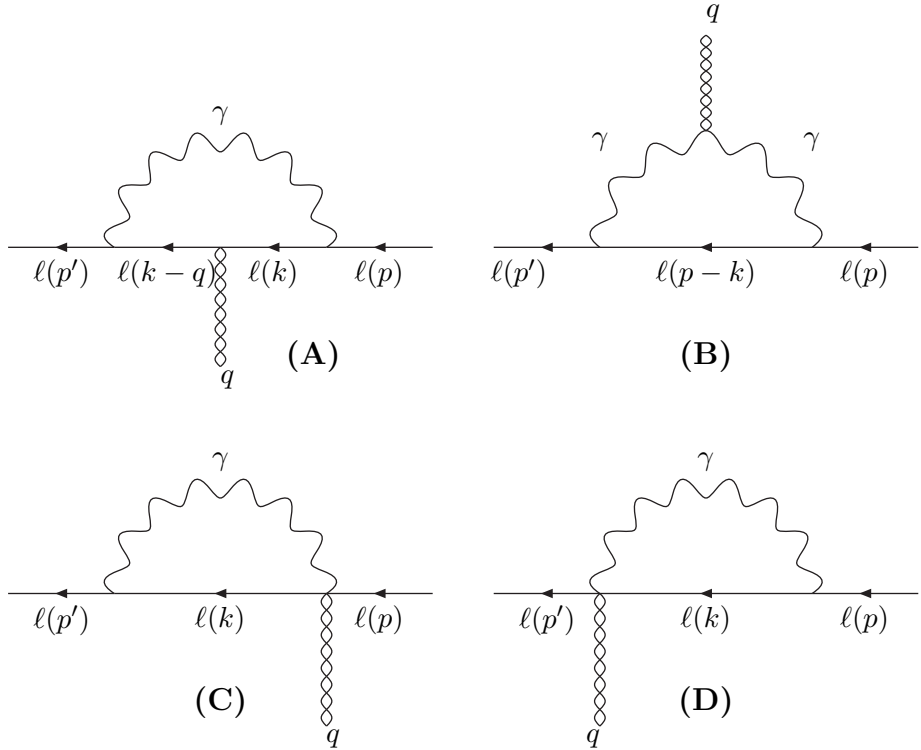

%
%
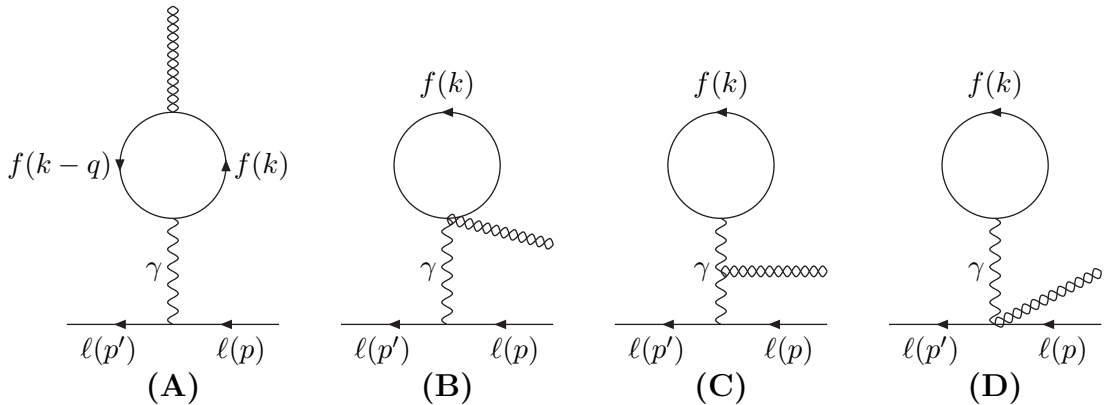
\begin{figure}[btp]
\begin{center}
%
%
\begin{picture}(100,170)(-50,-30)
\Text(0,-30)[cb]{\large\bf (A)}
\ArrowLine(40,0)(0,0)
\Text(35,-10)[cr]{$\ell(p)$}
\ArrowLine(0,0)(-40,0)
\Text(-35,-10)[cl]{$\ell(p')$}
\Photon(0,0)(0,40){2}{6}
\Text(-4,20)[r]{$\gamma$}
\ArrowArc(0,60)(20,90,270)
\ArrowArc(0,60)(20,-90,90)
\Text(23,60)[l]{$f(k)$}
\Text(-23,60)[r]{$f(k-q)$}
\Photon(0,80)(0,120){2}{6}
\Photon(0,80)(0,120){-2}{6}
\end{picture}
%
%
\begin{picture}(100,170)(-50,-30)
\Text(0,-30)[cb]{\large\bf (B)}
\ArrowLine(40,0)(0,0)
\Text(35,-10)[cr]{$\ell(p)$}
\ArrowLine(0,0)(-40,0)
\Text(-35,-10)[cl]{$\ell(p')$}
\Photon(0,0)(0,40){2}{6}
\Text(-4,20)[r]{$\gamma$}
\ArrowArc(0,60)(20,-90,270)
\Text(0,85)[b]{$f(k)$}
\Photon(0,40)(40,30){2}{6}
\Photon(0,40)(40,30){-2}{6}
\end{picture}
%
%
\begin{picture}(100,170)(-50,-30)
\Text(0,-30)[cb]{\large\bf (C)}
\ArrowLine(40,0)(0,0)
\Text(35,-10)[cr]{$\ell(p)$}
\ArrowLine(0,0)(-40,0)
\Text(-35,-10)[cl]{$\ell(p')$}
\Photon(0,0)(0,40){2}{6}
\Text(-4,20)[r]{$\gamma$}
\ArrowArc(0,60)(20,-90,270)
\Text(0,85)[b]{$f(k)$}
\Photon(0,20)(40,20){2}{6}
\Photon(0,20)(40,20){-2}{6}
\end{picture}
%
%
\begin{picture}(100,170)(-50,-30)
\Text(0,-30)[cb]{\large\bf (D)}
\ArrowLine(40,0)(0,0)
\Text(35,-10)[cr]{$\ell(p)$}
\ArrowLine(0,0)(-40,0)
\Text(-35,-10)[cl]{$\ell(p')$}
\Photon(0,0)(0,40){2}{6}
\Text(-4,20)[r]{$\gamma$}
\ArrowArc(0,60)(20,-90,270)
\Text(0,85)[b]{$f(k)$}
\Photon(0,0)(40,20){2}{6}
\Photon(0,0)(40,20){-2}{6}
\end{picture}
\end{center}
\caption[]{\sf Diagrams for the one-loop contribution to the
gravitational vertex of any charged lepton in a background of
electrons and nucleons.
\label{f:Ztype}}
\end{figure}
The various graviton couplings that are needed for the evaluation of
these diagrams have been given earlier. For completeness we summarize
here the relevant formulas.  For fermions, the Feynman rule for the
graviton-fermion-fermion vertex is $-i\kappa V_{\lambda\rho}(p,p')$,
where $V_{\lambda\rho}$ given in Eq.\ (\ref{Vmunu}), where $p$ and
$p'$ are the momenta of the incoming and the outgoing fermions
\cite{np:gravnu}.  The interaction involving the graviton, a photon
$A^\mu$ and a pair of charged fermions is represented by the Feynman
rule~\cite{np:photgrav,lingrav}  $ie\kappa
a_{\mu\nu\lambda\rho}\gamma^\nu$, where
\begin{eqnarray}
a_{\mu\nu\lambda\rho} = \eta_{\mu\nu} \eta_{\lambda\rho} -
\frac12 \left( \eta_{\mu\lambda} \eta_{\nu\rho} + \eta_{\nu\lambda}
\eta_{\mu\rho} \right) \,.
\label{a}
\end{eqnarray}
In addition, there is also a photon-photon-graviton vertex. For an
incoming photon $A_\mu(k)$ and an outgoing one $A^\nu(k')$, the
Feynman rule for this vertex is $-i\kappa
C_{\mu\nu\lambda\rho}(k,k')$, with~\cite{np:photgrav}
\begin{eqnarray}
C_{\mu\nu\lambda\rho} (k,k') &=& \eta_{\lambda\rho} ( \eta_{\mu\nu} k
\cdot k' - k'_\mu k_\nu ) - \eta_{\mu\nu} (k_\lambda k'_\rho +
k'_\lambda k_\rho ) \nonumber\\*
&& + k_\nu (\eta_{\lambda\mu} k'_\rho +
\eta_{\rho\mu} k'_\lambda)
+ k'_\mu (\eta_{\lambda\nu} k_\rho +
\eta_{\rho\nu} k_\lambda) \nonumber\\*
&& - k \cdot k' (\eta_{\lambda\mu} \eta_{\rho\nu} +
\eta_{\lambda\nu} \eta_{\rho\mu}) \,.
\label{C}
\end{eqnarray}
%

\subsection{Diagrams in Fig.\ \ref{f:Wtype}}
\subsubsection{Diagram \ref{f:Wtype}A}
The amplitude of the diagram in Fig.~\ref{f:Wtype}A can be written as 
\begin{eqnarray}
-i\kappa \Gamma^{(A)}_{\lambda\rho} (p,p') &=& \int {d^4k \over
(2\pi)^4} \; 
ie\gamma_\alpha \, iS_\ell(k') \, (-i\kappa) V_{\lambda\rho} (k,k') 
\, iS_\ell(k) \,
ie\gamma_\beta \,iD^{\alpha\beta} (k-p) \,,
\end{eqnarray}
where
\begin{eqnarray}
k' \equiv k-q \,.
\end{eqnarray}
As already explained, to determine the contribution to the dispersive
part of the vertex function we need to retain the terms that contain
the thermal part of only one of the propagators.  Any of them contains
some combination of the form
\begin{eqnarray}
\Lambda_{\lambda\rho} (k_1,k_2) \equiv \gamma_\alpha (\rlap/k_2+m_\ell)
V_{\lambda\rho} (k_1,k_2) (\rlap/k_1+m_\ell) \gamma^\alpha \,.
\end{eqnarray}
After some straightforward algebra, this can be written as 
\begin{eqnarray}
\Lambda_{\lambda\rho} (k_1,k_2) &=& -{1\over 2} 
\Big[ (k_1+k_2)_\rho (\rlap/k_1 \gamma_\lambda \rlap/k_2 +
m^2_\ell \gamma_\lambda) 
+ (k_1+k_2)_\lambda (\rlap/k_1 \gamma_\rho \rlap/k_2 + m^2_\ell
\gamma_\rho) \Big] \nonumber\\* 
&& + \eta_{\lambda\rho} \Big[ (k_1^2-m^2_\ell) (\rlap/k_2 -2m_\ell) + 
(k_2^2-m^2_\ell) (\rlap/k_1 - 2m_\ell) \Big] \nonumber\\* 
&& + 2m_\ell (k_1+k_2)_\lambda (k_1+k_2)_\rho \,.
\label{Lambda}
\end{eqnarray}

For the sake of convenience, we divide the total contribution into two
parts
\begin{eqnarray}
\label{Gammadecomp}
\Gamma'^{(A)}_{\lambda\rho} (p,p') = \Gamma'^{(A1)}_{\lambda\rho}
(p,p') + \Gamma'^{(A2)}_{\lambda\rho} (p,p') \,,
\end{eqnarray}
where $\Gamma'^{(A1)}_{\lambda\rho}$ contains the distribution
function of the photon and therefore contributes to the gravitational
vertex for all charged leptons, and $\Gamma'^{(A2)}_{\lambda\rho}$
contains the distribution function of the electrons and contributes
only to the vertex for the electrons. Changing the
integration variable from $k$ to $k+p$, we obtain
\begin{eqnarray}
\Gamma'^{(A1)}_{\lambda\rho} (p,p') &=& -e^2 \int {d^4k \over
(2\pi)^3} \; 
{\delta(k^2) \eta_\gamma(k) \over [(k+p')^2-m^2_\ell][(k+p)^2-m^2_\ell]}
\Lambda_{\lambda\rho} (k+p,k+p') \,,
\label{GamA1}
\end{eqnarray}
and similarly,
\begin{eqnarray}
\Gamma'^{(A2)}_{\lambda\rho} (p,p') &=& e^2 
\int {d^4k \over (2\pi)^3} \; \delta(k^2-m^2_\ell) \eta_\ell(k) 
\nonumber\\* && \times 
\Bigg( {\Lambda_{\lambda\rho} (k,k-q) 
\over [(k-q)^2-m_\ell^2](k-p)^2} +
{\Lambda_{\lambda\rho} (k+q,k) \over [(k+q)^2-m_\ell^2](k-p')^2}
\Bigg) \,.
\label{A2}
\end{eqnarray}
%

\subsubsection{Diagram \ref{f:Wtype}B}
For this diagram
\begin{eqnarray}
-i\kappa \Gamma^{(B)}_{\lambda\rho} (p,p') = \int {d^4k \over
(2\pi)^4} \; 
ie\gamma^\alpha \, iS_\ell(p-k) \, ie\gamma^\beta 
(-i\kappa) C_{\mu\nu\lambda\rho} (k,k') \, 
iD^{\nu\alpha} (k) iD^{\mu\beta} (k')\,,
\end{eqnarray}
and we decompose it in analogy with Eq.\ (\ref{Gammadecomp}).
The part that contains the photon distribution function is
\begin{eqnarray}
\Gamma'^{(B1)}_{\lambda\rho} (p,p')  
= e^2 \int {d^4k \over (2\pi)^4} \; 
\gamma^\nu S_{F\ell}(p-k) \gamma^\mu 
C_{\mu\nu\lambda\rho} (k,k') \, 
\Big[ \Delta_F (k) \Delta_T(k') + \Delta_F (k') \Delta_T(k) \Big] \,.
\end{eqnarray}
Making a change of the integration variable in one of the terms, this 
can be written as
\begin{eqnarray}
\Gamma'^{(B1)}_{\lambda\rho} (p,p') 
&=& e^2 \int {d^4k \over (2\pi)^3} \; \delta(k^2) \eta_\gamma(k) 
\nonumber\\* 
& \times & \Bigg[ { \gamma^\nu (\rlap/p'- \rlap/k + m_\ell) \gamma^\mu 
C_{\mu\nu\lambda\rho} (k+q,k) \over 
[(p'-k)^2-m_\ell^2] (k+q)^2}
+ {\gamma^\nu (\rlap/p- \rlap/k +m_\ell) \gamma^\mu 
C_{\mu\nu\lambda\rho} (k,k-q) \over [(p-k)^2-m_\ell^2] (k-q)^2} \Bigg]
\,, \quad 
\label{GamB1}
\end{eqnarray}
while
\begin{eqnarray}
\Gamma'^{(B2)}_{\lambda\rho} (p,p') &=& e^2 \int {d^4k \over (2\pi)^4} \; 
\gamma^\nu S_{T\ell}(k) \gamma^\mu 
C_{\mu\nu\lambda\rho} (p-k,p'-k) \, 
\Delta_F (p-k) \Delta_F (p'-k) 
\nonumber\\*
&=& -e^2 \int {d^4k \over (2\pi)^3} \; \delta(k^2-m_\ell^2) \eta_\ell(k) 
\gamma^\nu (\rlap/k + m_\ell) \gamma^\mu \; 
{C_{\mu\nu\lambda\rho} (p-k,p'-k) \over (p-k)^2 (p'-k)^2}
\label{vertex.B2}
\end{eqnarray}
gives the lepton background part.

\subsubsection{Diagrams \ref{f:Wtype}C and \ref{f:Wtype}D}
For these two diagrams the manipulations are similar and, omitting
the details, the results are
\begin{eqnarray}
\label{GammaC1D1}
\Gamma'^{(C1+D1)}_{\lambda\rho} (p,p') = -\, e^2
a_{\mu\nu\lambda\rho} 
\int {d^4k \over (2\pi)^3} \; 
\delta(k^2) \eta_\gamma(k) \left[ 
{\gamma^\mu (\rlap/k + \rlap/p' + m_\ell) \gamma^\nu \over(k+p')^2
-m_\ell^2} + 
{\gamma^\nu (\rlap/k + \rlap/p+m_\ell) \gamma^\mu \over (k+p)^2
-m_\ell^2} \right] \,,
\end{eqnarray}
and
\begin{eqnarray}
\Gamma'^{(C2+D2)}_{\lambda\rho} (p,p') = e^2 a_{\mu\nu\lambda\rho} 
\int {d^4k \over (2\pi)^3} \; 
\delta(k^2-m^2_\ell) \eta_\ell(k) \left[
{\gamma^\mu (\rlap/k +m_\ell) \gamma^\nu \over (k-p')^2} 
+ {\gamma^\nu (\rlap/k +m_\ell) \gamma^\mu \over (k-p)^2} 
\right] \,.
\end{eqnarray}
%

\subsection{Diagrams in Fig.~\ref{f:Ztype}}\label{s:Ztype}
\subsubsection{The question of the photon tadpole}\label{s:tadpole}
We are calculating the effective action given by the tree-level terms,
plus the $O(e^2)$ corrections that arise from the
diagrams in Figs.\ \ref{f:selfenergy}, \ref{f:Wtype} and
\ref{f:Ztype}.  Some of the diagrams contribute to the bilinear (or
kinetic) part of the effective action, from which we identify the
inertial mass and the wavefunction renormalization, while others
contribute to the interaction with the gravitational potential, from
which we identify the gravitational mass.

It is important to recall at this point that we are considering a
medium that is electrically neutral, which requires that the
parameters that characterize the composition of the medium be such
that the net contribution to the photon tadpole vanishes.
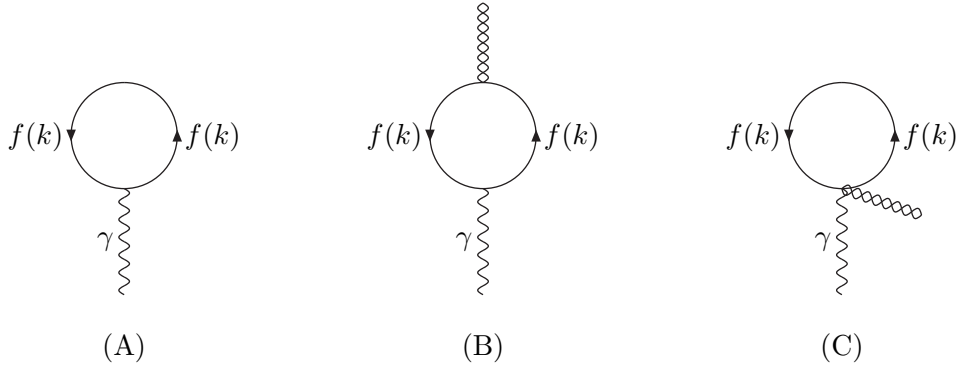
\begin{figure}[bhtp]
\begin{center}
\begin{picture}(100,140)(-50,-40)
\Photon(0,0)(0,40){2}{6}
\Text(0,-20)[c]{(A)}
\Text(-4,20)[r]{$\gamma$}
\ArrowArc(0,60)(20,90,270)
\ArrowArc(0,60)(20,-90,90)
\Text(23,60)[l]{$f(k)$}
\Text(-23,60)[r]{$f(k)$}
\end{picture}
\hspace{1cm}
\begin{picture}(100,140)(-50,-40)
\Text(0,-20)[c]{(B)}
\Photon(0,0)(0,40){2}{6}
\Text(-4,20)[r]{$\gamma$}
\ArrowArc(0,60)(20,90,270)
\ArrowArc(0,60)(20,-90,90)
\Text(23,60)[l]{$f(k)$}
\Text(-23,60)[r]{$f(k)$}
\Photon(0,80)(0,110){2}{4}
\Photon(0,80)(0,110){-2}{4}
\end{picture}
\hspace{1cm}
\begin{picture}(100,140)(-50,-40)
\Text(0,-20)[c]{(C)}
\Photon(0,0)(0,40){2}{6}
\Text(-4,20)[r]{$\gamma$}
\ArrowArc(0,60)(20,90,270)
\ArrowArc(0,60)(20,-90,90)
\Text(23,60)[l]{$f(k)$}
\Text(-23,60)[r]{$f(k)$}
\Photon(0,40)(30,30){2}{4}
\Photon(0,40)(30,30){-2}{4}
\end{picture}
\end{center}
\caption[]{\small\sf The one-loop diagrams 
that contribute to order $\kappa$ to the 
photon tadpole in a medium, in the presence of 
a static and homogeneous ($q = 0$) gravitational potential. The
fermion loop involves a sum over all the species of fermions present in
the medium.}\label{f:tadpole}
\end{figure}
The diagrams that contribute to the photon tadpole at the one-loop
level, in the presence of a static and homogeneous gravitational potential,
are shown in Fig.~\ref{f:tadpole}, where the graviton line represents
represents the $q=0$ background field. In the absence of the
background field, only the diagram \ref{f:tadpole}A
contributes to the photon tadpole. In that case, the
requirement that the tadpole vanishes yields the familiar condition
\begin{equation}
\label{neutralcond}
{\cal Q}^{(\ref{f:tadpole}A)} \equiv 
\sum_f {\cal Q}_f\left[2\int\frac{d^3K}{(2\pi)^3}[f_f(E_K)-f_{\bar f}(E_K)]
\right]
= 0
\end{equation}
where $f_{f}$ and $f_{\bar f}$ are given by Eq.\ (\ref{distfunctions}), and
the sum is over all species of fermions in the
medium, the charge of each species being denoted by ${\cal Q}_f$ with
the convention that ${\cal Q}_e = -1$.
In this case, the quantity ${\cal Q}^{(\ref{f:tadpole}A)}$ is identified
with the total charge of the medium.
However, in the presence of the background field, and
to the order that we are calculating, we have to take into account
the contributions of the diagrams \ref{f:tadpole}B and \ref{f:tadpole}C
to the photon tadpole or, equivalently, to 
the total charge of the system. If we denote them
by ${\cal Q}^{(\ref{f:tadpole}B)}$ and ${\cal Q}^{(\ref{f:tadpole}C)}$
respectively, it is the sum 
${\cal Q}^{(\ref{f:tadpole}A)} + {\cal Q}^{(\ref{f:tadpole}B)}
+ {\cal Q}^{(\ref{f:tadpole}C)}$ that must be zero for
the photon tadpole to vanish.
Physically, this means
that the number density of the particles are not determined by their free
distribution functions. The particle distributions 
rearrange themselves in a way that depends on
the background gravitational field.

This has the following implication for our calculation.  
Firstly, the
unadorned tadpole of Fig. \ref{f:tadpole}A is now itself of order $\kappa$
because of the charge neutrality condition. Since the
diagrams \ref{f:Ztype}C and \ref{f:Ztype}D contain an explicit factor of
$\kappa$ apart from the unadorned tadpole, their contribution is
actually of order $\kappa^2$ and therefore we can neglect them. 
Secondly, the diagram
shown in Fig. \ref{f:selfenergy}B cancels the $q$-independent
contributions from the diagrams \ref{f:Ztype}A and
\ref{f:Ztype}B. Since the loop in diagram \ref{f:Ztype}B in
independent of $q$, this diagram is totally canceled.

In summary, the only contribution from the diagrams shown in
Fig.~\ref{f:Ztype} arises from the $q^2$-dependent part of the tadpole
subdiagram of Fig.~\ref{f:Ztype}A. When multiplied by the
photon propagator, it gives zero for the $\delta(q^2)$ part
in the propagator while its linear term in $q^2$ cancels the $1/q^2$ 
in the other part.
This latter contribution will be labeled by the letter `X' in
order not to confuse it with the contributions of
Fig.~\ref{f:Wtype}A.

\subsubsection{The non-vanishing contribution}\label{s:non0}
We denote the vertex contribution coming from Fig.~\ref{f:Ztype}A by 
\begin{eqnarray}
\Gamma^{(X)}_{\lambda\rho} (p,p') &=& 
{e^2 \gamma^\alpha \over q^2} X_{\lambda\rho\alpha} (q) \,,
\label{Xq}
\end{eqnarray}
where $X_{\lambda\rho\alpha} (q)$ is the photon-graviton mixing
diagram with external momentum $q$
\begin{eqnarray}
X_{\lambda\rho\alpha} (q) = 
\sum_f \int {d^4k \over (2\pi)^4} {\rm Tr} \Big[
V_{\lambda\rho} (k,k') iS_f(k) i{\cal Q}_f \gamma_\alpha iS_f(k')
\Big] \,. 
\label{defX}
\end{eqnarray}
Then, taking the above discussion above into account, the quantity
which will appear in the expression for the gravitational mass is
given by 
\begin{eqnarray}
\widetilde\Gamma^{(X)}_{\lambda\rho} (p,p') &=& 
{e^2 \gamma^\alpha \over q^2} \Big[ X_{\lambda\rho\alpha} (q) 
- X_{\lambda\rho\alpha} (0) \Big] \,.
\label{Xq-X0}
\end{eqnarray}
As already mentioned, the sum in Eq.\ (\ref{defX}) 
is over all species of fermions in the
medium, the charge of each species being denoted by ${\cal Q}_f$ with
the convention that ${\cal Q}_e = -1$.  
The medium-dependent
contribution to $X_{\lambda\rho\alpha} (q)$ can be written as
\begin{eqnarray}
X_{\lambda\rho\alpha} (q) 
= \sum_f {\cal Q}_f
\int {d^4k\over (2\pi)^3} 
\; \delta(k^2-m_f^2) \eta_f(k) \left[ 
{A_{\lambda\rho\alpha}(k,k-q) \over q^2 - 2k\cdot q} + 
{A_{\lambda\rho\alpha}(k+q,k) \over q^2 + 2k\cdot q}  \right] \,,
\label{GamX}
\end{eqnarray}
where, for arbitrary 4-momenta $k_1$ and $k_2$,
\begin{eqnarray}
A_{\lambda\rho\alpha} (k_1,k_2) &=& 
{\rm Tr}  \Big[
V_{\lambda\rho} (k_1,k_2) 
(\rlap/k_1 +m_f) \gamma_\alpha (\rlap/k_2+m_f) \Big]
\nonumber\\* 
&=& \Big[ 
(2k_{1\lambda} k_{1\rho} + k_{1\lambda} k_{2\rho} + k_{2\lambda}
k_{1\rho}) k_{2\alpha} 
+ (m_f^2-k_1\cdot k_2) (\eta_{\lambda\alpha} k_{1\rho} +
\eta_{\rho\alpha} k_{1\lambda}) \nonumber\\*
&& - 2 \eta_{\lambda\rho} (k_1^2-m_f^2)
k_{2\alpha} \Big] 
+ \Big[ k_1 \leftrightarrow k_2\Big] \,.
\end{eqnarray}
Putting $k^2=m_f^2$, we obtain
\begin{eqnarray}
A_{\lambda\rho\alpha} (k,k-q) &=& 
[8k_\lambda k_\rho - 4 (k_\lambda
q_\rho + k_\rho q_\lambda) + 2 q_\lambda q_\rho] k_\alpha 
- [4k_\lambda k_\rho - (k_\lambda
q_\rho + k_\rho q_\lambda)] q_\alpha \nonumber\\* && 
+ k\cdot q [\eta_{\lambda\alpha} (2k-q)_\rho + \eta_{\rho\alpha}
(2k-q)_\lambda ] 
- 2 \eta_{\lambda\rho} (q^2-2k\cdot q) k_\alpha \,.
\label{A...}
\end{eqnarray}
Since $A_{\lambda\rho\alpha} (k_1,k_2) = A_{\lambda\rho\alpha}
(k_2,k_1)$ by definition, $A_{\lambda\rho\alpha} (k+q,k)$ is
obtained by changing the sign of $q$ in this expression.

Finally, we mention that the complete one-loop vertex function
satisfies the transversality condition, which is implied by the gravitational
gauge invariance of the theory.
This is shown in Appendix~\ref{app:transv}.

\section{Calculation of the gravitational mass}
\label{s:calc}
As seen in Eq.\ (\ref{operationalM'}), there are three types of
$O(e^2)$ correction to the gravitational mass.  One of them is
proportional to the inertial mass that was calculated in
Sec.~\ref{s:SE}, and another one involves the wave function
renormalization factor derived in Sec.~\ref{s:Z}. In this section we
find the contributions from the one-loop vertex diagrams. Since the
expressions for those already have an explicit factor of $e^2$ outside
the integral, to evaluate them we can use the tree-level values for
the dispersion relation and the spinors associated with the external
lepton.

\subsection{Terms with the photon distribution from Fig.~\ref{f:Wtype}}
We first evaluate those terms obtained in Sec.~\ref{s:indgrav}
that contain the photon distribution function. In fact, if the
temperature of the ambient medium is low ($T\ll m_e$) and the chemical
potential of the background electrons is zero, these are the only
terms that contribute and they are precisely the ones that were
calculated in Ref.\ \cite{DHR84}. 
Since we have performed the calculations in a different
way, using 1-particle irreducible diagrams only, the following results
serve as a good checkpoint between the earlier calculations of
Ref.\ \cite{DHR84} and ours.

\subsubsection{Contribution (A1)}
{}From the formula for the gravitational mass given in 
Eq.\ (\ref{operationalM'}), 
it follows that we need to calculate the 
vertex only for $p=p'$, in which case
\begin{eqnarray}
\Gamma'^{(A1)}_{\lambda\rho} (p,p) &=& -e^2 \int {d^4k \over
(2\pi)^3} \; 
{\delta(k^2) \eta_\gamma(k) \over 4(k\cdot p)^2}
\Lambda_{\lambda\rho} (k+p,k+p) \,.
\end{eqnarray}
{}From Eq.\ (\ref{Lambda}) it follows that, for any 4-vector $y^\mu$,
\begin{eqnarray}
\Lambda_{\lambda\rho} (y,y) = -4 y_\lambda y_\rho (\rlap/y - 2m_\ell) +
(y^2-m_\ell^2) \bigg[ (\gamma_\lambda y_\rho + \gamma_\rho y_\lambda) +
2\eta_{\lambda\rho} (\rlap/y - 2m_\ell) \bigg] \,,
\label{Lambdayy}
\end{eqnarray}
which leads to
\begin{eqnarray}
\overline u_s(p)\Gamma'^{(A1)}_{\lambda\rho} (p,p)u_s(p) & = &
-  e^2 \int {d^4k \over
(2\pi)^3} \; 
{\delta(k^2) \eta_\gamma(k) \over (k\cdot p)^2}\nonumber\\*
&\times& \overline u_s(p)\left[
-{k\cdot p\over m_\ell} k_\lambda p_\rho
+ m_\ell (k_\lambda k_\rho + p_\lambda p_\rho) 
+ {(k\cdot p)^2 \over m_\ell} \eta_{\lambda\rho}\right]u_s(p) \,,
\end{eqnarray}
where we have used Eq.\ (\ref{ugu}) and omitted the terms odd in
$k$, which integrate to zero.  Using the notation
\begin{eqnarray}
m'_{(A1)} = 
(2v^\lambda v^\rho - \eta^{\lambda\rho}) \left[
\overline u_s(p) \Gamma'^{(A1)}_{\lambda\rho}(p,p) 
u_s(p)\right]_{p^\mu = (m_\ell,\vec 0)}
\,,
\end{eqnarray}
we obtain
\begin{eqnarray}
m'_{(A1)} &=& e^2 \int {d^3K \over
(2\pi)^3} \; f_\gamma(K) \bigg[ {1\over m_\ell K} - {m_\ell\over K^3} \bigg]
\nonumber\\* 
&=& {e^2T^2 \over 12m_\ell} - {e^2 m_\ell \over 2\pi^2} \int_0^\infty {dK\over
K} \; f_\gamma(K) \,.
\label{A1final}
\end{eqnarray}
The remaining integral is infrared divergent, but its contribution
to the gravitational mass is canceled
by a similar term that arises from the wavefunction renormalization,
as we show below.

\subsubsection{Contribution (B1)}
This term has to be treated carefully because the denominators
in the integrand of Eq.\ (\ref{GamB1}) 
vanish for $q=0$. However, a careful evaluation of this term,
following the procedure indicated in 
Eq.\ (\ref{operationalM'}),
shows that the limit exists. 
Denoting
\begin{eqnarray}
m'_{(B1)} \equiv 
(2v^\lambda v^\rho - \eta^{\lambda\rho})
\lim_{\vec P\rightarrow 0}\left\{\left[
\vphantom{\frac{1}{2}}
\overline u_s(p')\Gamma'^{(B1)}_{\lambda\rho}(p,p')u_s(p)
\right]_\lplim\right\} \,,
\end{eqnarray}
the result is
\begin{eqnarray}
m'_{(B1)} &=& - {e^2T^2 \over 3m_\ell} \,.
\label{B1final}
\end{eqnarray}
The details of the derivation of this result are given in
Appendix~\ref{app:tough}. 

\subsubsection{Contributions (C1+D1)}
We can proceed as in the evaluation of $m'_{(A1)}$ above. Thus, from
Eq.\ (\ref{GammaC1D1}),
\begin{eqnarray}
\overline u_s(p)\Gamma'^{(C1+D1)}_{\lambda\rho} (p,p)u_s(p) = 
-\, e^2 a_{\mu\nu\lambda\rho} 
\int {d^4k \over (2\pi)^3} \; 
\frac{\delta(k^2) \eta_\gamma(k)}{k\cdot p}
\left[\overline u_s(p)\gamma^\mu \rlap/k \gamma^\nu u_s(p)\right] \,,
\end{eqnarray}
using the fact that $a_{\mu\nu\lambda\rho}$ is symmetric in the
indices $\mu,\nu$. Then using 
\begin{eqnarray}
(2v^\lambda v^\rho - \eta^{\lambda\rho}) a_{\mu\nu\lambda\rho} =
-\eta_{\mu\nu} - 2v_\mu v_\nu 
\end{eqnarray}
it follows that
\begin{eqnarray}
\label{C1D1final}
m'_{(C1+D1)} & \equiv  &
(2v^\lambda v^\rho - \eta^{\lambda\rho}) 
\left[\overline u_s(p)\Gamma'^{(C1+D1)}_{\lambda\rho}(p,p) 
u_s(p)\right]_{p^\mu = (m_\ell,\vec 0)} \nonumber\\
& = & 0 \,.
\end{eqnarray} 
%

\subsection{Terms with the electron distribution from Fig.~\ref{f:Wtype}}
These terms contribute only to the vertex involving electrons and
positrons.  The integration over $k_0$ and the angular variables can
be done exactly.  The remaining integral can be evaluated analytically
only for special cases of the distribution functions, some of which we
consider afterwards.

\subsubsection{Contribution (A2)}
As can be seen from Eq.\ (\ref{A2}), 
the denominators of the integrand of this term vanish as
$q\to0$. Consequently, the prescription indicated in 
Eq.\ (\ref{operationalM'}) has to be followed carefully in this case.
As we show in detail in Appendix \ref{app:tough:A2}, 
defining
\begin{eqnarray}
m'_{(A2)} \equiv 
(2v^\lambda v^\rho - \eta^{\lambda\rho})
\lim_{\vec P\rightarrow 0}\left\{\left[
\vphantom{\frac{1}{2}}
\overline u_s(p')\Gamma'^{(A2)}_{\lambda\rho}(p,p')u_s(p)
\right]_\lplim\right\} \,,
\end{eqnarray}
the final result for this term is
\begin{eqnarray}
m'_{(A2)} & = & {e^2 \over m_e} \int {d^3K \over (2\pi)^32E_K} 
\left\{ 
{2E_K^2-m_e^2 \over E_K} \left( {E_K-2m_e \over E_K-m_e} 
\; {\partial f_e \over
\partial E_K} + {E_K+2m_e \over E_K+m_e} \; {\partial f_{\bar e} \over
\partial E_K} \right)\right. \nonumber\\* 
&& + {2E_K^4 - E_K^3m_e - 5E_K^2m_e^2 + 2E_Km_e^3 - 2m_e^4 
\over m_e E_K^2
(E_K-m_e)} f_e \nonumber\\* 
&& - \left.{2E_K^4 + E_K^3m_e - 5E_K^2m_e^2 - 2E_Km_e^3 - 2m_e^4 
\over m_e E_K^2
(E_K+m_e)} f_{\bar e}  
\right\} \,,
\label{A2final}
\end{eqnarray}
where $E_K$ is defined in Eq.\ (\ref{EK}).

\subsubsection{Contribution (B2)}
{}From Eq.\ (\ref{vertex.B2}) it is seen that
the integrand is not singular in the limit $q\to0$.
Therefore we can evaluate directly
\begin{eqnarray}
\Gamma'^{(B2)}_{\lambda\rho} (p,p) 
&=& -e^2 \int {d^4k \over (2\pi)^3} \; \delta(k^2-m_e^2) \eta_e(k) 
\gamma^\nu (\rlap/k+m_e) \gamma^\mu \; 
{C_{\mu\nu\lambda\rho} (p-k,p-k) \over (p-k)^4} \,,
\end{eqnarray}
and the contribution to the gravitational mass is given by
\begin{eqnarray}
m'_{(B2)} = 
(2v^\lambda v^\rho - \eta^{\lambda\rho}) 
\left[\overline u_s(p)\Gamma'^{(B2)}_{\lambda\rho}(p,p) u_s(p)\right]_{
p^\mu = (m_e,\vec 0)} \,.
\end{eqnarray}
In the expression for $C_{\mu\nu\lambda\rho}$, any term having a
factor of $(p-k)_\mu$ or $(p-k)_\nu$ does not contribute to the
integral. This is because, within the spinors, we can write
\begin{eqnarray}
\gamma^\nu (\rlap/k+m_e) \gamma^\mu (p-k)_\mu = 
\gamma^\nu (\rlap/k+m_e) (m_e-\rlap/k) 
= \gamma^\nu (m_e^2-k^2) \,,
\end{eqnarray}
which vanishes because of the $\delta$-function. The argument is
similar for $(p-k)_\nu$. Thus,
\begin{eqnarray}
&&\hspace{-1cm}
(2v^\lambda v^\rho - \eta^{\lambda\rho}) \overline u_s(p) 
\gamma^\nu (\rlap/k+m_e)\gamma^\mu C_{\mu\nu\lambda\rho}(p-k,p-k)
u_s(p) \nonumber\\
&&=
\overline u_s(p)\Big\{
8(\rlap/k-2m_e)  (p\cdot v - k\cdot v)^2 + 4 (m_e - 2k\cdot v \rlap/v)
(p-k)^2 \Big\} u_s(p)\,, \quad
\end{eqnarray}
ignoring all terms which have a factor of $k^2-m_e^2$.
Using Eqs.\ (\ref{ugu}) and (\ref{unorm}), we then obtain
\begin{eqnarray}
m'_{(B2)}
&=& -\, {2e^2\over m_e^2} \int {d^4k \over
(2\pi)^3} \; \delta(k^2-m_e^2) \eta_e(k) \left( k_0 + {m_e^2 
\over k_0-m_e}
\right) 
\nonumber\\ 
&=& -\, {e^2\over m_e^2} \int {d^3K\over
(2\pi)^3} \left( f_e - f_{\bar e} \right)  
- {e^2\over 2\pi^2} \int dK \; {K^2 \over E_K}
\left[ {f_e(E_K) \over E_K -m_e } - {f_{\bar e}(E_K) 
\over E_K +m_e }\right] .
\label{B2final}
\end{eqnarray}
%

\subsubsection{Contributions (C2+D2)}
Similarly, for this term we can evaluate directly
\begin{eqnarray}
m'_{(C2 + D2)} = 
(2v^\lambda v^\rho - \eta^{\lambda\rho}) 
\left[\overline u_s(p)\Gamma'^{(C2 + D2)}_{\lambda\rho}(p,p) u_s(p)\right]_{
p^\mu = (m_e,\vec 0)} \,,
\end{eqnarray}
with
\begin{eqnarray}
\Gamma'^{(C2+D2)}_{\lambda\rho} (p,p) = 2e^2 a_{\mu\nu\lambda\rho} 
\int {d^4k\over
(2\pi)^3} \delta(k^2-m_e^2) \eta_e(k) \; {\gamma^\mu (\rlap/k+m_e)
\gamma^\nu \over (k-p)^2} \,.
\end{eqnarray}
By straight forward algebra
\begin{eqnarray}
(2v^\lambda v^\rho - \eta^{\lambda\rho}) a_{\mu\nu\lambda\rho}
\gamma^\mu (\rlap/k+m_e) \gamma^\nu = 4 \rlap/k - 6m_e - 4 k\cdot v
\rlap/v \,,
\end{eqnarray}
and using Eqs.\ (\ref{ugu}) and (\ref{unorm}),
\begin{eqnarray}
m'_{(C2+D2)} &=& 6e^2 \int {d^4k \over
(2\pi)^3} \; \delta(k^2-m_e^2) \eta_e(k) {1 \over k_0-m_e}
\nonumber\\ 
&=& {3e^2\over 2\pi^2} \int dK  \; {K^2 \over E_K} 
\Bigg[ {f_e(E_K) \over E_K-m_e} - {f_{\bar e} (E_K) \over 
E_K+m_e} \Bigg] .
\label{C2D2final}
\end{eqnarray}
%

\subsection{Terms from Fig.~\ref{f:Ztype}}
The contribution to the gravitational mass due to this term is
\begin{eqnarray}
\label{mXdef}
m'_{(X)} \equiv 
(2v^\lambda v^\rho - \eta^{\lambda\rho})
\lim_{\vec P\rightarrow 0}\left\{\left[
\vphantom{\frac{1}{2}}
\overline u_s(p') \widetilde\Gamma^{(X)}_{\lambda\rho}(p,p') u_s(p)
\right]_\Qlim\right\} \,,
\end{eqnarray}
where, from Eq.\ (\ref{Xq-X0}),
\begin{eqnarray}
\label{XQ-X0}
\widetilde\Gamma^{(X)}_{\lambda\rho} (p,p') \Bigg|_{q^0=0}  = 
- \, {e^2 \gamma^\alpha \over Q^2} 
\left[ X_{\lambda\rho\alpha} (q) \Big|_{q^0 = 0} 
- X_{\lambda\rho\alpha} (0) \right] \,.
\end{eqnarray}
Using the expression for $A_{\lambda\rho\alpha}$ from Eq.\ (\ref{A...})
we obtain
\begin{eqnarray}
(2v^\lambda v^\rho - \eta^{\lambda\rho})
X_{\lambda\rho\alpha} (q) \Big|_{q^0 = 0} & = &  
8\sum_f {\cal Q}_f
\int {d^4k\over (2\pi)^3} 
\; \delta(k^2-m_f^2) \eta_f(k) k_0 v_\alpha \nonumber\\*
& \times&
\left[ 
(2k_0^2-m_f^2 - \frac12 Q^2) 
\left( {1\over 2 \vec K\cdot \vec Q - Q^2} -
{1\over 2 \vec K\cdot \vec Q + Q^2} \right)\right] \,,
\end{eqnarray}
where we have omitted the terms that vanish by symmetric integration
over $\vec K$, as well as all those terms that are independent 
of $q$, because they drop out of 
Eq.\ (\ref{XQ-X0}), and in addition
all the terms that are
proportional to $q_\alpha$, because in Eq.\ (\ref{mXdef}) they yield
a factor of $\rlap/q$ which vanishes between spinors.
Performing the integration over $k^0$, 
\begin{eqnarray}
(2v^\lambda v^\rho - \eta^{\lambda\rho})
X_{\lambda\rho\alpha} (q) \Big|_{q^0 = 0}
&=& 
4v_\alpha\sum_f {\cal Q}_f
\int {d^3K\over (2\pi)^3} 
\; \Big( f_f - f_{\bar f} \Big)
\nonumber\\* & \times & 
(2E_K^2-m_f^2 - \frac12 Q^2) 
\left( {1\over 2 \vec K\cdot \vec Q - Q^2} -
{1\over 2 \vec K\cdot \vec Q + Q^2} \right).\;
\label{GamX.4}
\end{eqnarray}
For the term that contains an explicit factor of $Q^2$ in the
numerator we use the angular integration formula of Eq.\
(\ref{angint}), which yields
\begin{eqnarray}
(2v^\lambda v^\rho - \eta^{\lambda\rho})
X_{\lambda\rho\alpha} (q)\Big|_{q^0 = 0}
\!\!&= & v_\alpha\sum_f {\cal Q}_f \Bigg[ 
{Q^2\over 2\pi^2} 
\int dK \; \Big( f_f - f_{\bar f} \Big)
\nonumber\\* 
& +& \!4 \int \! {d^3K\over (2\pi)^3} 
\; \Big( f_f - f_{\bar f} \Big)
(2E_K^2-m_f^2) 
\left( {1\over 2 \vec K\cdot \vec Q - Q^2} -
{1\over 2 \vec K\cdot \vec Q + Q^2} \right) \!\!\Bigg] .
\nonumber\\* 
\label{GamX.5}
\end{eqnarray}
The evaluation of the rest of the integral is presented in 
Appendix \ref{app:tough:X}. Substituting the results into 
Eq.\ (\ref{XQ-X0}), the contribution of this diagram to the
gravitational mass is found to be given by
\begin{eqnarray}
m'_{(X)} 
\label{Xfinal}
&=& - e^2 \sum_f {{\cal Q}_f \over 6\pi^2} 
\int dK \Bigg[  \Big( f_f - f_{\bar f} \Big) - 
 {2E_K^2-m_f^2\over 2E_K} {\partial \over \partial E_K} \Big( f_f -
f_{\bar f} \Big) \Bigg] \,.
\end{eqnarray}
%

\subsection{Summary}
Starting from Eq.\ (\ref{operationalM'}), the total contribution to
the gravitational mass of charged leptons can be written in the form
\begin{eqnarray}
M'_\ell = m_\ell + m'_{\ell1} + m'_{\ell2} + m'_{(X)} \,,
\label{Mell'def}
\end{eqnarray}
where $m'_{(X)}$ is the contribution from Eq.\ (\ref{Xfinal}), which
is the same for all charged leptons, $m'_{\ell1}$ represents the terms
that contain the photon distribution function, and $m'_{\ell2}$
contains the terms that depend on the electron distribution
function. They are given as follows.

Substituting into Eq.\ (\ref{operationalM'}) the results given in
Eqs.\ (\ref{A1final}), (\ref{B1final}) and (\ref{C1D1final}), and
using the expression for the wave-function normalization and the
correction to the inertial mass given in Eqs.\ (\ref{Zpsi1}) and 
(\ref{m1}), we find
\begin{eqnarray}
m'_{\ell1} = - {e^2T^2 \over 12m_\ell} \,,
\end{eqnarray}
in agreement with the DHR result\cite{DHR84}, quoted in Eq.\
(\ref{DHR:Mg}).  Notice that the infrared divergence contained in the
$m'_{(A1)}$ term cancels with a similar one that arises from the wave
function renormalization correction $\zeta_{\ell 1}$.

The terms from the diagrams in Fig.~\ref{f:Wtype} that involve the
fermion distribution function contribute only to the gravitational
mass of the electron, and therefore
\begin{eqnarray}
m'_{\mu2} = m'_{\tau2} = 0 \,.
\end{eqnarray}
The individual contributions of this type to the electron
gravitational mass appear in Eqs.\ (\ref{A2final}), (\ref{B2final})
and (\ref{C2D2final}). 
Substituting those results into Eq.\ (\ref{operationalM'}),
and using the results for the
inertial mass and the wave-function normalization factor, given in
Eqs.\ (\ref{me2}) and (\ref{ZU2}) respectively, we obtain
\begin{eqnarray}
m'_{e2} & = & {\displaystyle e^2 \over \displaystyle\pi^2 m_e}
\int_0^\infty dK \; {K^2  \over 2E_K} \Bigg\{ 
\left( {3\over 2} + {m_e^2 \over E_K^2} - {m_e \over
E_K - m_e} \right) f_e(E_K) \nonumber\\*
& & + \left( {3\over 2} + {m_e^2 \over E_K^2} + {m_e \over
E_K + m_e} \right) f_{\bar e} (E_K) \nonumber\\* 
& & +
{2E_K^2-m_e^2 \over 2E_K} \bigg( {E_K-2m_e \over E_K-m_e} \; 
{\partial f_e \over
\partial E_K} + {E_K+2m_e \over E_K+m_e} \; {\partial f_{\bar e} \over
\partial E_K} \bigg) \Bigg\} \,.
\label{m'2summary}
\end{eqnarray}

The corresponding formulas for the antileptons are obtained by making
the substitution $v^\mu \rightarrow -v^\mu$, as indicated in
Eq. (\ref{CoperationalMbar'}).  Since the dependence of
$\Gamma'_{\lambda\rho}$ on $v^\mu$ arises only through the factors
$\eta_f$ and $\eta_\gamma$ defined in Eqs.\ (\ref{etaf}) and
(\ref{etab}), the substitution $v^\mu \rightarrow -v^\mu$ is
equivalent to the prescription
\begin{eqnarray}
\label{antiprescription1}
\mu_f \rightarrow -\mu_f \,,
\end{eqnarray}
or equivalently,
\begin{eqnarray}
\label{antiprescription2}
f_f \leftrightarrow f_{\bar f} \,.
\end{eqnarray}
Thus,
\begin{eqnarray}
\label{Mbarell'def}
M'_{\bar\ell} = m_\ell + m'_{\ell1} + m'_{{\bar\ell}2} - m'_{X} \,,
\end{eqnarray}
where
\begin{eqnarray}
m'_{\bar\mu2} = m'_{\bar\tau2} = 0\,,
\end{eqnarray}
while the result for $m'_{{\bar e}2}$ is obtained from 
Eq.\ (\ref{m'2summary}}) by making the substitution
$f_e \leftrightarrow f_{\bar e}$.

\section{Results for particular cases}
\label{s:cases}
In contrast with the $m_{\ell 1}$ and $m'_{\ell 1}$ terms,
which depend on the photon momentum distribution, 
$m_{e2}$, $m_{\bar e2}$, $m'_{e2}$, $m'_{\bar e2}$ and $m'_{(X)}$ depend
on the fermion distribution functions and cannot be
evaluated exactly in the general case.
Therefore, for illustration, we consider in detail 
their calculation for the specific
situation in which the background is composed of non-relativistic protons 
and electrons. In this case we can set (for $f = e,p$)
\begin{eqnarray}
f_{\bar f}(E) \approx 0
\end{eqnarray}
and
\begin{eqnarray}
\label{Enonrel}
E_K \approx m_f + \frac{K^2}{2m_f} \,.
\end{eqnarray}
We consider in detail two cases separately,
according to whether the electron gas is classical or degenerate.

\subsection{Classical electron gas and classical proton gas}
In this situation we can set
\begin{eqnarray}
\label{classdist}
f_f (E) = e^{-\beta(E-\mu_f)}
\end{eqnarray}
for both $f = e,p$. This implies the relation
\begin{eqnarray}
\label{classrel}
{\partial f_f\over \partial E} = -\beta f_f \,,
\end{eqnarray}
as well as the integration formula
\begin{eqnarray}
\int dK \; K^{2r} f_f = 2 \pi^{3/2} \Gamma \left(r+\frac12 \right) 
\left( {\beta\over 2m_f} \right)^{1-r} n_f\,,
\label{Kint}
\end{eqnarray}
where $n_f$ is the number density, given by
\begin{eqnarray}
n_f = 2 \int {d^3K \over (2\pi)^3} \; f_f(E) \approx 2 \left( {m_f
\over 2\pi\beta} \right)^{3/2} e^{-\beta(m_f-\mu_f)} \,.
\end{eqnarray}

Let us consider $m_{e2}$ and $m_{\bar e2}$, given in Eqs.\ (\ref{me2})
and (\ref{mbare2}), respectively. Setting $f_{\bar e} = 0$ and using
Eq.\ (\ref{Enonrel}) to expand the co-efficients of $f_e$ in the
integrands in powers of $K$, the remaining integrals are evaluated by
means of Eq.\ (\ref{Kint}) to yield
\begin{eqnarray}
\label{me2mbare2class}
m_{e2} & = & - \frac{e^2n_e}{2m_e T} + O(n_e/m_e^2)\,,
\nonumber\\
m_{\bar e2} & = & \frac{3e^2n_e}{8m_e^2} + 
O(n_e T^2/m_e^4)\,.
\end{eqnarray}
Similarly, from Eq.\ (\ref{Xfinal}) we obtain for this case
\begin{eqnarray}
\label{Xclassical}
m'_{(X)} & = & - {7e^2 \over 24T} \sum_{f = e,p} {{\cal Q}_f n_f \over
m_f} 
+ O(n_e T/m_e^3)\nonumber\\*
& \approx & \frac{7e^2n_e}{24m_e T} \,,
\end{eqnarray}
where we have used the charge-neutrality condition which, neglecting
terms $O(\kappa$), is simply $\sum_f{\cal Q}_f n_f=0$.  Applying the
same procedure in Eq.\ (\ref{m'2summary}), the leading contribution,
in powers of $T/m_e$, comes from the $\partial f_e/\partial E_K$ term
in that equation, and leads to
\begin{eqnarray}
m'_{e2} = {e^2 n_e \over 2T^2} + O(n_e/Tm_e) \,.
\end{eqnarray}
By the substitution indicated in Eq.\ (\ref{antiprescription2}),
the corresponding result for the positron is
\begin{eqnarray}
m'_{\bar e2} = - {3e^2 n_e \over 8m_eT} + O(n_e/m^2_e)\,.
\end{eqnarray}
Therefore, using Eqs.\ (\ref{Melldef}), (\ref{Mell'def}) and
(\ref{Mbarell'def}), the inertial and gravitational masses for charged
leptons $\ell$ other than the electron are obtained as
\begin{eqnarray}
M_{\bar\ell} = M_{\ell} & = & m_\ell + {e^2 T^2 \over 12m_\ell} \,, 
\nonumber\\*
M'_{\ell,\bar\ell} &=& m_\ell - {e^2 T^2 \over 12m_\ell} \pm
\frac{7e^2n_e}{24 m_e T}\,,
\end{eqnarray}
where the upper sign corresponds to the leptons and the lower one to
the anti-leptons. The corresponding formulas for the electron are
\begin{eqnarray}
M_{e} & = & m_e + {e^2 T^2 \over 12m_e} - {e^2 n_e \over 2m_eT}\,,
\nonumber\\*
M'_e & = & m_e - {e^2 T^2 \over 12m_e} + {e^2 n_e \over 2T^2} \,,
\end{eqnarray}
and for the positron they are
\begin{eqnarray}
M_{\bar e} &=& m_e + {e^2 T^2 \over 12m_e} + {3e^2 n_e \over 8m_e^2}  \,,
\nonumber\\* 
M'_{\bar e} &=& m_e - {e^2 T^2 \over 12m_e} - {2e^2 n_e \over 3m_eT} \,.
\end{eqnarray}

We now estimate how large these corrections could be for the
electron. Those due to the photon background
were estimated by DHR\cite{DHR84} and were found to be extremely
small. Therefore, neglecting that contribution,
the fractional change in the inertial and gravitational mass
is given by
\begin{eqnarray}
\label{fracMe}
\left| {M_e - m_e \over m_e} \right| & = & {e^2 n_e \over 2m_e^2T}\\*
\label{fracM'e}
\left| {M'_e - m_e \over m_e} \right| & = & {e^2 n_e \over 2m_eT^2} \,.
\end{eqnarray}
Although it may seem that the effects are more noticeable as the
temperature decreases, they are bounded by the condition that the
electron gas is non-degenerate and non-interacting,
which requires that~\cite{ll:pk}
\begin{eqnarray}
T > \frac{n_e^{2/3}}{m_e}
\label{classicalcond}
\end{eqnarray}
and
\begin{eqnarray}
\label{freecond}
T > {e^2 \over r_{\rm av}} \sim e^2 n_e^{1/3} \,,
\end{eqnarray}
since $r_{\rm av} \sim n_e^{-1/3}$.  Using the fact that $n_p=n_e$,
it follows that the corresponding conditions for the proton gas do not
imply further restrictions, because they are automatically satisfied
whenever Eqs.\ ({\ref{classicalcond}) and (\ref{freecond}) hold.

By writing the right-hand side of Eq.\ (\ref{fracMe}) in
the alternative forms 
\begin{eqnarray}
\frac{e^2 n_e}{2m^2_eT} & = & \left(\frac{e^2 n^{1/3}_e}{T}\right)
\left(\frac{n^{2/3}_e}{2m^2_e}\right) \nonumber\\
& = & \left(\frac{n_e^{2/3}}{m_e T}\right)^2
\left(\frac{T}{2m_e}\right)\left(\frac{e^2m_e}{n_e^{1/3}}\right)\,,
\end{eqnarray}
it is seen that
\begin{eqnarray}
\left| {M_e - m_e \over m_e} \right| < \left\{
\begin{array}{ll}
e^4/2 & \mbox{if $n_e^{1/3} < e^2 m_e$} \,, \\[12pt]
T/2m_e & \mbox{if $n_e^{1/3} > e^2 m_e$}\,.
\end{array}
\right.
\end{eqnarray}
Similarly, writing 
\begin{eqnarray}
\frac{e^2n_e}{2m_eT^2} & = & \frac{1}{2}\left(\frac{n^{1/3}_e}{e^2m_e}\right)
\left(\frac{e^2 n^{1/3}_e}{T}\right)^2 \nonumber\\
& = & \frac{1}{2}\left(\frac{n_e^{2/3}}{m_e T}\right)^2
\left(\frac{e^2 m_e}{n_e^{1/3}}\right) \,,
\end{eqnarray}
it follows that
\begin{eqnarray}
\left| {M'_e - m_e \over m_e} \right| <  \frac{1}{2}
\end{eqnarray}
in either case.
Therefore, while the fractional correction to the electron's
inertial mass is likely to be small in most situations 
with the conditions that we are presently considering,
the fractional change in the gravitational mass 
could be substantial.  For example, if we use the temperature and
density at the solar core, i.e., $T=1.57\times10^7$\,K,
$n_e=9.5\times10^{25}$\,cm$^{-3}$, we obtain
\begin{eqnarray}
\left| {M_e - m_e \over m_e} \right| &=& 9.8 \times 10^{-5}\,,
\nonumber\\ 
\left| {M'_e - m_e \over m_e} \right| &=& 3.5 \times 10^{-2}\,,
\end{eqnarray}
which shows that the correction to the gravitational mass of the
electron can at least be appreciable in realistic physical
situations.

\subsection{Degenerate electron gas and classical proton gas}
In this case
\begin{eqnarray}
T \ll \frac{n_e^{2/3}}{m_e} \sim \frac{K_F^2}{m_e} \,,
\label{degenerate}
\end{eqnarray}
where $K_F$ is the Fermi momentum of the electron gas. We assume
\begin{eqnarray}
K_F \ll m_e
\label{NRcondition}
\end{eqnarray}
so that the electrons are non-relativistic, and
\begin{eqnarray}
\label{Freecondition}
K_F > e^2 m_e \,,
\end{eqnarray}
which implies that the average kinetic energy of an electron is larger
than the average Coulomb interaction energy $\sim e^2 n_e^{1/3} \sim
e^2 K_F$, and therefore the electron gas can be treated as an ideal
gas.  Under these conditions, the protons can be treated as a weakly
coupled Boltzmann gas if we assume that the weak coupling
condition
\begin{eqnarray}
T \gg e^2 n_p^{1/3} \sim e^2 K_F
\label{pfreecond}
\end{eqnarray}
is satisfied\cite{raffelt:book}. 
Remembering that $K_F \ll m_e \ll m_p$, this in turn
implies the non-degeneracy condition
\begin{eqnarray}
T \gg \frac{K_F^2}{m_p} \sim \frac{n_p^{2/3}}{m_p} \,.
\label{pNDcond}
\end{eqnarray}
Therefore, Eq.\ (\ref{classdist})
applies to the proton, while for the electron
\begin{eqnarray}
\label{fedegen}
f_e = \Theta (K_F-K)
\end{eqnarray}
with
\begin{eqnarray}
K_F = (3\pi^2 n_e)^{1/3} \,,
\end{eqnarray}
which in turn imply the relation
\begin{eqnarray}
\label{derfedegen}
{df_e \over dK} = -\delta (K_F-K)\,.
\end{eqnarray}

We repeat the calculation of the quantities $m_{e2}$, $m_{\bar e2}$,
$m'_{e2}$, $m'_{\bar e2}$ and $m'_{X}$ for this case, neglecting the
terms that are a factor $\sim O(K^2_F/m^2_e)$ smaller than the ones
that we retain. {}From Eqs.\ (\ref{me2}) and (\ref{mbare2}), setting
$f_{\bar e} = 0$ and using Eq.\ (\ref{Enonrel}), we obtain
\begin{eqnarray}
\label{me2degen}
m_{e2} & = & - \frac{e^2K_F}{2\pi^2} \,,\nonumber\\
m_{\bar e2} & = & \frac{e^2 K_F^3}{8\pi^2 m_e^2} \,.
\end{eqnarray}
{}From Eq.\ (\ref{Xfinal}),
\begin{eqnarray}
m'_{(X)} = \frac{e^2}{6\pi^2} 
\int dK \Bigg[ f_e - 
\left(K + \frac{m^2_e}{2K}\right) 
\frac{df_e}{dK}\Bigg] - {7e^2 \over 24T} {n_p \over m_p} \,,
\end{eqnarray}
where we have borrowed the result for the proton contribution from 
Eq.\ (\ref{Xclassical}), while in the electron term we have
expressed $E_K$ in terms of $K$ and used 
\begin{eqnarray}
{d \over dE_K} = \frac{E_K}{K}\frac{d}{dK}
\label{EdE=KdK}
\end{eqnarray}
for any function of $E_K$.  Using Eqs.\ (\ref{fedegen}) and
(\ref{derfedegen}) this finally yields
\begin{eqnarray}
\label{Xdegen}
m'_{(X)} = \frac{e^2 m^2_e}{12\pi^2 K_F} \,. 
\end{eqnarray}
Here we have neglected the proton contribution because it is $\sim
e^2K_F^3/(Tm_p) \ll e^2K_F$ from Eq.\ (\ref{pNDcond}).  In a similar
fashion, from Eq.\ (\ref{m'2summary}),
\begin{eqnarray}
m'_{e2} = \frac{e^2 m^2_e}{2\pi^2 K_F} \,,
\end{eqnarray}
and by the substitution indicated in Eq.\ (\ref{antiprescription2}),
the corresponding result for the positron is
\begin{eqnarray}
m'_{\bar e2} = - \frac{3e^2 K_F}{8\pi^2} \,.
\end{eqnarray}

Thus, substituting these results into Eqs.\ (\ref{Melldef}),
(\ref{Mell'def}) and (\ref{Mbarell'def}), we obtain the following
expressions for the inertial and gravitational masses, retaining only
the leading terms in powers of $K_F/m_e$.  For the charged leptons
$\ell$ other than the electron,
\begin{eqnarray}
\label{Melldegen}
M_{\bar\ell} = M_\ell & = & m_\ell + {e^2 T^2 \over 12m_\ell}
\nonumber\\*
M'_{\ell,\bar\ell} & = & m_\ell - {e^2 T^2 \over 12m_\ell} \pm 
\frac{e^2 m^2_e}{12\pi^2 K_F} \,,
\end{eqnarray}
with the upper sign corresponding to the leptons and the lower one to
the anti-leptons, while for the electron
\begin{eqnarray}
\label{Medegen}
M_e &=& m_e + {e^2 T^2 \over 12m_e} - {e^2 K_F\over
2\pi^2}  \,, \nonumber\\* 
M'_e &=& m_e - {e^2 T^2 \over 12m_e} 
+ \frac{7e^2 m^2_e}{12\pi^2 K_F} \,,
\end{eqnarray}
and for the positron
\begin{eqnarray}
\label{Mbaredegen}
M_{\bar e} &=& m_e + {e^2 T^2 \over 12m_e} + {e^2 K_F^3\over
8\pi^2m_e^2}  \,, \nonumber\\* 
M'_{\bar e} &=& m_e - {e^2 T^2 \over 12m_e} - \frac{e^2 m_e^2}{12\pi^2 K_F} \,.
\end{eqnarray}

It is interesting to note that Eqs.\ (\ref{degenerate}) and
(\ref{NRcondition}) imply that the photon contributions in Eqs.\
(\ref{Melldegen})-(\ref{Mbaredegen}) are much smaller than the
contribution due to the electron background in each case. In fact,
using Eq.\ (\ref{Freecondition}), we see that the fractinal
corrections to the gravitational mass can be as large as about
$7/12\pi^2$ for the electron and $1/12\pi^2$ for the positron and the
other leptons.

\section{Conclusions}\label{s:conclu}
In this work we derived a general operational formula that
expresses the gravitational mass of a fermion in terms of the
gravitational vertex function.  Using that formula as the staring point, we
have studied the $O(e^2)$ corrections to the gravitational
interactions of a charged lepton in the presence of a matter
background.  This calculation extends and complements previous
calculations along similar lines, in various useful ways.

{}From a technical point of view, the calculations that we have
presented have employed various finite-temperature-field-theory
techniques that can be useful also in other contexts. For example, we
have shown in detail how a careful treatment of the wavefunction
renormalization factor, which arises from considering the one-particle
reducible diagrams in the proper way, is instrumental in the
cancelation of an infrared divergent contribution that arises from the
photon contribution to the proper vertex function.

On the other hand, a well known problem that arises in this type of
calculation is the ambiguity of the finite temperature Green functions
when they are evaluated at zero momentum\cite{zeromomprob}.  This
property is usually due to the fact that the different mathematical
limits correspond to different physical situations, so in those cases
the resolution of the apparent paradox lies in recognizing the
appropriate correspondence with the physical situation at hand
\cite{np:thermalse}.  The calculations that we have presented have
illustrated this in a particularly convincing way.  The operational
formula for the gravitational mass given, in Eq.\ (\ref{operationalM'}),
indicate the precise order in which the various limits must be taken,
according to the physical situation that we considered.  As we have
shown, the careful application of that prescription has allowed us to
evaluate all the integrals involved, in a unique and well-defined way,
including those that superficially seem to be singular, without having
to introduce by hand any special regularization technique.

The calculations and the results are also important from a
phenomenological point of view. As we have indicated, in a matter
background with a non-zero chemical potential such as the Sun, the
matter contributions to the gravitational mass are proportional to the
electron and nucleon densities and its magnitude can be appreciable.
These matter contributions dominate over the photon-background
contribution, especially in those situations in which $T \ll m_e$, for
which the photon contribution becomes negligible. Moreover, the
matter-induced corrections to the gravitational mass are different for
the various charged lepton flavors, and are not be the same for the
corresponding antiparticles.  There are situations in which mass
differences, intrinsic or induced, have important physical
implications, such as the neutron-proton mass difference in the
context of the nucleosynthesis calculations in the Early Universe.
Although our work has focused in the case of the charged leptons,
similar considerations can be applied to the other fermions as well.
The question of the possible implications of this type of mass
correction in specific situations is an important one, but is outside
the scope of the present work. Nevertheless, our calculations have
provided a necessary ingredient for being able to consider them in a
systematic manner, and set the stage for their further study on a firm
basis.

\paragraph*{Acknowledgment}
This work has been  partially 
supported (JFN) by the U.S. National Science Foundation Grant PHY-9900766. 
\appendix

\leftline{\null\hrulefill\null}\nopagebreak
\section*{Appendices}
\section{Transversality of the vertex}\label{app:transv}
It is useful to verify that the complete vertex function to $O(e^2)$,
obtained in Sec.~\ref{s:indgrav}, satisfies the transversality
conditions
\begin{eqnarray}
q^\lambda \overline U_s(p') \, \Gamma_{\lambda\rho}(p,p') \, U_s(p)
= q^\rho \overline U_s(p') \, \Gamma_{\lambda\rho}(p,p') \, U_s(p) =
0
\end{eqnarray}
to this order. Since the vertex is symmetric in the Lorentz indices
$\lambda,\rho$, either of these relations guarantees the other.
In order to simplify the notation, in the remainder of this appendix
we omit the subscript $s$ in the spinors.

In order to verify this relation,
the important point is that we must
include all the terms upto $O(e^2)$. Since the one-loop
terms in the induced vertex are already
$O(e^2)$, for them we can adopt the tree-level definition of
the spinors, i.e.
\begin{eqnarray}
\rlap/p \, u(p) = mu(p)\,, 
\qquad
\overline u(p') \rlap/p' = m \overline u(p)\,, 
\label{treespinors}
\end{eqnarray}
as well as the tree-level on-shell conditions 
\begin{eqnarray}
p^2=p'^2=m^2 \,. 
\label{treedisp}
\end{eqnarray}
In the appendices, we use the tree level mass $m$ without any
subscript, implying $m_\ell$, $m_e$ or $m_f$ which should be
understood from the context.  Also note that the photon distribution
function as well as the associated $\delta$-function are even in $k$,
and therefore those terms which are odd in $k$ in the rest 
of the integrand do not contribute.

We first show that the vertex contribution from
Fig.~\ref{f:Ztype}A is transverse by itself. From Eq.\ (\ref{A...}),
\begin{eqnarray}
q^\lambda A_{\lambda\rho\alpha}(k,k-q) = - (q^2-2k\cdot q) (4k_\rho
k_\alpha + k\cdot q \eta_{\rho\alpha}) \,,
\end{eqnarray}
where we omit the terms that are proportional to $q_\alpha$ 
because, in Eq.\ (\ref{Xq}),
they will yield $\rlap/q$ which vanishes between the spinors.
Changing the sign of $q$ in the last equation yields
\begin{eqnarray}
q^\lambda A_{\lambda\rho\alpha}(k+q,k) = (q^2+2k\cdot q) (4k_\rho
k_\alpha - k\cdot q \eta_{\rho\alpha}) \,,
\end{eqnarray}
and as a result $q^\lambda \overline u(p')
\Gamma^{(X)}_{\lambda\rho}(p,p') u(p)$ turns out to be
proportional to $\sum_f Q_f (n_f - n_{\bar f})$, which is zero
to this order.

As for the other diagrams, straightforward algebra gives the
following results:
\begin{eqnarray}
q^\lambda \overline u(p') \Gamma'^{(A1)}_{\lambda\rho} (p,p') u(p) 
&=& -\, {e^2\over 4} \int {d^4k \over (2\pi)^3} \;
\delta(k^2) \eta_\gamma(k) \nonumber\\*
&\times & \overline u(p') 
\Bigg[ {4m k_\rho - (p+5p')_\rho \rlap/k + 2 k\cdot p 
\gamma_\rho \over k\cdot p'} - \Big( p \leftrightarrow p' \Big)
\Bigg] u(p) \nonumber\\
q^\lambda \overline u(p') \Gamma'^{(A2)}_{\lambda\rho} (p,p') u(p) 
&=& {e^2 \over 4} \int {d^4k \over
(2\pi)^3} \; \delta(k^2-m^2) \eta_F(k) \nonumber\\*
&\times & \overline u(p') \Bigg[ 
{4(\rlap/k-2m) k_\rho + \rlap/k (p+p')_\rho - 2 k \cdot p' \gamma_\rho
\over m^2- k\cdot p} 
- \Big( p \leftrightarrow p' \Big)
\Bigg] u(p) \,, \nonumber\\
q^\lambda \overline u(p') \Gamma'^{(B1)}_{\lambda\rho} (p,p')  u(p) 
&=& e^2 \int {d^4k \over (2\pi)^3} \;
\delta(k^2) \eta_\gamma(k) \nonumber\\*
&\times& \overline u(p') \Bigg[ {m k_\rho - \rlap/k p'_\rho \over
k\cdot p'} - {m k_\rho - \rlap/k p_\rho \over k\cdot p}
\Bigg] u(p) \nonumber\\
q^\lambda \overline u(p') \Gamma'^{(B2)}_{\lambda\rho} (p,p')  u(p) 
&=& -\, e^2 \int {d^4k \over (2\pi)^3} \; \delta(k^2-m^2)
\eta_F(k) \nonumber\\* 
&\times & \overline u(p') 
\left[ {(\rlap/k - 2m) k_\rho + mp_\rho \over m^2 - k\cdot p} -
{(\rlap/k - 2m) k_\rho + mp'_\rho \over m^2 - k\cdot p'} \right]
u(p) \,, \nonumber\\ 
q^\lambda \overline u(p') \Gamma'^{(C1+D1)}_{\lambda\rho} (p,p') u(p)
&=& {e^2 \over 2} \int {d^4k \over (2\pi)^3} \;
\delta(k^2) \eta_\gamma(k) \left( {1 \over k\cdot p'} + {1 \over
k\cdot p} \right) \nonumber\\*
&\times& 
\overline u(p') \Big[  \rlap/k q_\rho + k\cdot q
\gamma_\rho \Big] u(p) \,, \nonumber\\
q^\lambda \overline u(p') \Gamma'^{(C2+D2)}_{\lambda\rho} (p,p') u(p) 
&=& -\, {e^2 \over 2} \int {d^4k \over (2\pi)^3} \; \delta(k^2-m^2)
\eta_F(k) \left(
{1\over m^2-k\cdot p'} + {1\over m^2-k\cdot p}
\right) \nonumber\\* 
&\times& \overline u(p') \Big[ (\rlap/k-3m) q_\rho + k\cdot q
\gamma_\rho \Big] u(p)  \,. 
\end{eqnarray}
Therefore, 
adding all the one-loop contributions to the vertex, we obtain
\begin{eqnarray}
q^\lambda \overline u(p') \Gamma'^{(1)}_{\lambda\rho} (p,p') u(p) 
&=& {e^2 \over 4} \int {d^4k \over (2\pi)^3} \;
\delta(k^2) \eta_\gamma(k) \nonumber\\*
&\times& 
\overline u(p') \Bigg[ {\rlap/k \over k\cdot p'} \left( 3p_\rho -
p'_\rho \right)  - \Big( p \leftrightarrow p' \Big)
\Bigg] u(p) \,,
\label{qGamma1} \\
q^\lambda \overline u(p') \Gamma'^{(2)}_{\lambda\rho} (p,p') u(p) 
&=& {e^2 \over 4} \int {d^4k \over (2\pi)^3} \;
\delta(k^2-m^2) \eta_F(k) \nonumber\\*
&\times& 
\overline u(p') \Bigg[  
{(\rlap/k -2m)(3p'_\rho - p_\rho) - 2 k\cdot p \gamma_\rho \over m^2 -
k\cdot p } - 
\Big( p \leftrightarrow p' \Big)
\Bigg] u(p) \,.
\label{qGamma2}
\end{eqnarray}

We need to add to these the tree-level contribution to the gravitational
vertex that appears in Eq.\ (\ref{Vmunu}). In this case, we must
include the $O(e^2)$ corrections to the equation for the spinors, which
arise from the self-energy diagrams of Sec.~\ref{s:SE}. Thus, for
this part, using Eq.\ (\ref{diraceq}) and its hermitian conjugate
\begin{eqnarray}
\overline U(p') \Big( \rlap/p' - m - \Sigma(p') \Big) = 0 \,,
\label{Up'}
\end{eqnarray}
we obtain
\begin{eqnarray}
\overline U(p') \rlap/q  U(p) = \overline
U(p') \Big( \Sigma'(p) - \Sigma'(p') \Big) U(p) \,,
\end{eqnarray}
which in turn yields
\begin{eqnarray}
q^\lambda \overline U(p') \, V_{\lambda\rho}(p,p') \, U(p) =
\frac14 \overline 
U(p') \bigg[ (3p'-p)_\rho \Sigma'(p) 
+ p^2 \gamma_\rho 
- \Big( p \leftrightarrow p' \Big) \bigg]  U(p) \,.
\label{qV}
\end{eqnarray}
This can be cast in a different form by multiplying Eq.\
(\ref{diraceq}) from the left by $\overline
U(p')\gamma_\rho(\rlap/p+m)$ and Eq.\ (\ref{Up'}) from the right
by $(\rlap/p'+m)\gamma_\rho U(p)$ and taking the difference of the
resulting equations. This gives
\begin{eqnarray}
(p^2-p'^2) \; \overline U(p') \gamma_\rho U(p) = \overline U(p') \Big[
\gamma_\rho (\rlap/p+m) \Sigma'(p) - \Sigma'(p') (\rlap/p'+m)
\gamma_\rho \Big] U(p) \,,
\label{psq-p'sq}
\end{eqnarray}
and substituting this result into Eq.\ (\ref{qV}), we obtain
\begin{eqnarray}
q^\lambda \overline U(p') \, V_{\lambda\rho}(p,p') \, U(p) &=&
\frac14 \overline 
U(p') \bigg[ (3p'-p)_\rho \Sigma'(p) - (3p-p')_\rho \Sigma'(p')
\nonumber\\* 
&& + \gamma_\rho (\rlap/p+m) \Sigma'(p) - \Sigma'(p') (\rlap/p'+m)
\gamma_\rho   \bigg]  U(p) \,.
\label{qVfinal}
\end{eqnarray}
Since $\Sigma'$ is explicitly of $O(e^2)$
while we are interested in results to $O(e^2)$ only,
we can use the tree-level spinors on the right-hand side.
Using Eq.\ (\ref{treedisp}) in Eq.\ (\ref{Sigma1}),
we can write the self-energy contribution involving the photon
distribution function as
\begin{eqnarray}
\Sigma'_1(p) &=& e^2 \int {d^4k \over (2\pi)^3} \; \delta(k^2)
\eta_\gamma(k) \; {\rlap/ k \over k\cdot p}  \,,
\label{Sigma1short}
\end{eqnarray}
disregarding terms odd in $k$. Similarly, from Eq.\
(\ref{Sigma2}), the part containing the Fermi distribution function
can be written as
\begin{eqnarray}
\Sigma'_2(p) 
&=& -\, e^2 \int {d^4k \over (2\pi)^3} \; 
\delta(k^2-m^2) \eta_F(k) \;
{\rlap/ k - 2m \over m^2 - k\cdot p} \,.
\label{Sigma2short}
\end{eqnarray}
Substituting these forms into Eq.\ (\ref{qVfinal}) and using the 
identities
\begin{eqnarray}
\overline u(p') \gamma_\rho (\rlap/p+m) \rlap/k u(p) 
&=& 2 k\cdot p \; \overline u(p') \gamma_\rho u(p) \nonumber\\* 
\overline u(p') \rlap/k (\rlap/p'+m) \gamma_\rho u(p) 
&=& 2 k\cdot p' \; \overline u(p') \gamma_\rho u(p) \,,
\end{eqnarray}
we see that Eq.\ (\ref{qVfinal}) cancels the contribution from the
loop diagrams given in Eqs.\ (\ref{qGamma1}) and (\ref{qGamma2}) to
this order. This proves the transversality of the
effective vertex.

\section{Apparently singular contributions}\label{app:tough}
\subsection{The B1 contribution}\label{app:tough:B1}
We start from the formula given in Eq.\ (\ref{GamB1}), from which it
follows that
\begin{eqnarray}
\label{GamB1a}
\Gamma'^{(B1)}_{\lambda\rho} (p,p) & = & 
-\, \frac{e^2}{2} 
\lim_{\vec Q\rightarrow 0}
\int\frac{d^4k}{(2\pi)^3}\delta(k^2) \eta_\gamma(k) \nonumber\\
& \times & \left[
\frac{\gamma^\nu (\rlap/p- \rlap/k +m) \gamma^\mu 
C_{\mu\nu\lambda\rho} (k,k - q)}{k\cdot p(2\vec K\cdot\vec Q - Q^2)} -
\frac{\gamma^\nu(\rlap/{p} - \rlap/{k} - \rlap/{q} + m) 
\gamma^\mu C_{\mu\nu\lambda\rho} (k + q,k)}{(k\cdot p + \vec K\cdot\vec Q)
(2\vec K\cdot\vec Q + Q^2)} \right] \,,
\end{eqnarray}
where we have put
\begin{eqnarray}
q^\mu = (0,\vec Q) \qquad k^\mu = (k^0,\vec K) \,.
\label{q&k}
\end{eqnarray}
In order to take the limit $\vec Q\rightarrow 0$, our strategy is to
expand the coefficients of the factors \hbox{$1/(2\vec K\cdot\vec Q
\pm Q^2)$} in powers of $\vec Q$. Of the resulting terms in the
coefficients, those which are quadratic in $\vec Q$ do not contribute
in the $\vec Q\rightarrow 0$ limit and therefore we need to keep only
the terms that are at most linear in $\vec Q$. 

Using the property $C_{\mu\nu\lambda\rho}(k + q,k) =
C_{\nu\mu\lambda\rho}(k,k + q)$, we can write
\begin{eqnarray}
\label{Cexp}
C_{\mu\nu\lambda\rho} (k,k - q) & = & C_{\mu\nu\lambda\rho} (k,k) - 
C'_{\mu\nu\lambda\rho} (k,q) \nonumber\\
C_{\mu\nu\lambda\rho} (k + q,k) & = & C_{\mu\nu\lambda\rho} (k,k) +
C'_{\nu\mu\lambda\rho} (k,q) \,,
\end{eqnarray}
where
\begin{eqnarray}
C'_{\mu\nu\lambda\rho} (k,q) & = &
\eta_{\lambda\rho}(\eta_{\mu\nu}k\cdot q - q_\mu k_\nu)
- \eta_{\mu\nu}(k_\lambda q_\rho + q_\lambda k_\rho) \nonumber\\* 
&& +
k_\nu(\eta_{\lambda\mu}q_\rho + \eta_{\rho\mu}q_\lambda) +
q_\mu(\eta_{\lambda\nu}k_\rho + \eta_{\rho\nu}k_\lambda) -
k\cdot q(\eta_{\lambda\mu}\eta_{\rho\nu} +
\eta_{\lambda\nu}\eta_{\rho\mu}) \,.
\end{eqnarray}
To first order in $Q$, we can also put
\begin{eqnarray}
\frac{1}{k\cdot p + \vec K\cdot\vec Q} = \frac{1}{k\cdot p} - 
\frac{\vec K\cdot\vec Q}{(k\cdot p)^2} \,.
\end{eqnarray}
This enables us to decompose $\Gamma'^{(B1)}_{\lambda\rho} (p,p)$ in
the following four terms:
\begin{eqnarray}
\Gamma'^{(B1a)}_{\lambda\rho}(p) \!\!\!& = & -\, \frac{e^2}{2} 
\lim_{\vec Q\rightarrow 0}
\int {d^4k \over (2\pi)^3} \; \delta(k^2) \eta_\gamma(k) 
{C_{\mu\nu\lambda\rho}(k,k)
\gamma^\nu (\rlap/p- \rlap/k +m) \gamma^\mu \over k\cdot p}
\nonumber\\* 
&& \times \left[\frac{1}{2\vec K\cdot\vec Q - Q^2} -
\frac{1}{2\vec K\cdot\vec Q + Q^2}\right] \nonumber\\[12pt]
\Gamma'^{(B1b)}_{\lambda\rho}(p) & = & -\, \frac{e^2}{2} 
\lim_{\vec Q\rightarrow 0}
\int\frac{d^4k}{(2\pi)^3}\delta(k^2) \eta_\gamma(k)
{C_{\mu\nu\lambda\rho}(k,k) \gamma^\nu\rlap/{q}\gamma^\mu \over 
k\cdot p(2\vec K\cdot Q + Q^2)} \nonumber\\[12pt] 
\Gamma'^{(B1c)}_{\lambda\rho}(p) & = & -\, \frac{e^2}{2} 
\lim_{\vec Q\rightarrow 0}
\int {d^4k \over (2\pi)^3} \; \delta(k^2) \eta_\gamma(k) 
{C_{\mu\nu\lambda\rho}(k,k) \gamma^\nu (\rlap/p- \rlap/k +m) \gamma^\mu 
\over (k\cdot p)^2}
\left[\frac{\vec K\cdot\vec Q}{2\vec K\cdot\vec Q + Q^2}\right]
\nonumber\\[12pt]
\Gamma'^{(B1d)}_{\lambda\rho}(p) & = & -\, \frac{e^2}{2} 
\lim_{\vec Q\rightarrow 0}
\int\frac{d^4k}{(2\pi)^3}\delta(k^2) \eta_\gamma(k)
{ \gamma^\nu(\rlap/p - \rlap/k + m)\gamma^\mu
\over k\cdot p} \nonumber\\*
&&  \times\left[
-\frac{C'_{\mu\nu\lambda\rho}(k,q)}{2\vec K\cdot \vec Q - Q^2}
-\frac{C'_{\nu\mu\lambda\rho}(k,q)}{2\vec K\cdot \vec Q + Q^2}
\right] . 
\end{eqnarray}
We carry out these integrals one by one.

Eliminating the manifestly $k$-odd terms from the integrand and
performing the $k_0$-integra\-tion, we obtain
\begin{eqnarray}
\Gamma'^{(B1a)}_{\lambda\rho}(p) 
= e^2 \lim_{\vec Q\rightarrow 0}
\int\frac{d^3K}{(2\pi)^3 2K} f_\gamma(K) 
{4k_\lambda k_\rho \rlap/k \over k\cdot p}
\left[\frac{1}{2\vec K\cdot\vec Q - Q^2} -
\frac{1}{2\vec K\cdot\vec Q + Q^2}\right] \,,
\end{eqnarray}
using $k^2=0$. The expression within the square brackets is finite for
$Q\to0$.  Therefore, in the spinors we can set $p=p'$, and using Eq.\
(\ref{ugu}) we then obtain
\begin{eqnarray}
m'_{(B1a)} = {4e^2 \over m} \lim_{\vec Q\rightarrow 0}
\int\frac{d^3K}{(2\pi)^3} f_\gamma(K) K
\left[\frac{1}{2\vec K\cdot\vec Q - Q^2} -
\frac{1}{2\vec K\cdot\vec Q + Q^2}\right] \,. 
\label{m'B1a}
\end{eqnarray}
We can perform the integration over the angular variables in $\vec K$,
the integral being understood, as usual, in terms of the principal
value part. That gives
\begin{eqnarray}
\int d\Omega 
\frac{1}{2\vec K \cdot \vec Q - Q^2} = 
- \int d\Omega 
\frac{1}{2\vec K \cdot \vec Q + Q^2} = 
-\, \frac{\pi}{K^2} + {\cal O}(Q^2) \,,
\label{angint}
\end{eqnarray}
so that
\begin{eqnarray}
m'_{(B1a)} = - \, {e^2 \over \pi^2 m} 
\int dK \; f_\gamma(K) K 
& = & -\, \frac{e^2T^2}{6m} \,.
\label{B1Aresult}
\end{eqnarray}

As for the next contribution, it is straightforward to verify that
\begin{eqnarray}
(2v^\lambda v^\rho - \eta^{\lambda\rho})\gamma^\nu\rlap/{q}\gamma^\mu
C_{\mu\nu\lambda\rho}(k,k) = 
4\vec K\cdot \vec Q (\rlap/k - 2k\cdot v\rlap/v )\,,
\end{eqnarray}
using $q\cdot v=q_0=0$. So
\begin{eqnarray}
(2v^\lambda v^\rho - \eta^{\lambda\rho})
\Gamma'^{(B1b)}_{\lambda\rho}(p) 
& = & e^2 \int\frac{d^4k}{(2\pi)^3}\delta(k^2) \eta_\gamma(k)
\frac{(-\rlap/k + 2k\cdot v\rlap/v)}
{k\cdot p} \,.
\end{eqnarray}
Now using Eqs.\ (\ref{ugu}) and (\ref{unorm}), carrying out the
integral over $k^0$, and finally putting $P=0$, we get
\begin{eqnarray}
\label{B1Bresult}
m'_{(B1b)} = 
{2e^2 \over m} \int\frac{d^3K}{(2\pi)^3 2K}f_\gamma(K) 
= \frac{e^2 T^2}{12m} \,.
\end{eqnarray}

Similarly,
\begin{eqnarray}
\Gamma'^{(B1c)}_{\lambda\rho}(p) 
& = & -\, \frac{e^2}{2}
\int {d^3K \over (2\pi)^3 2K} \; f_\gamma(K) 
\gamma^\nu (\rlap/p + m) \gamma^\mu C_{\mu\nu\lambda\rho}(k,k)
\frac{1}{(k\cdot p)^2} \,,
\end{eqnarray}
and
\begin{eqnarray}
(2v^\lambda v^\rho -
\eta^{\lambda\rho})\Gamma'^{(B1c)}_{\lambda\rho}(p) &=&  
-{2e^2\over m} \int {d^3K \over (2\pi)^3 2K} \; f_\gamma(K)
\frac{1}{(k\cdot p)^2} 
\nonumber\\* 
&& \times \left[
-(k\cdot p)^2 - 2m^2(k\cdot v)^2 + 4(k\cdot p)(k\cdot v)(p\cdot v)
\right]
\end{eqnarray}
so that
\begin{eqnarray}
\label{B1Cresult}
m'_{(B1c)} = 
-{2e^2\over m} \int {d^3K \over (2\pi)^3 2K} \; f_\gamma(K) 
= -\frac{e^2 T^2}{12m} \,.
\end{eqnarray}

For $\Gamma'^{(B1d)}_{\lambda\rho}$ we first perform the integral over
$k^0$. Remembering that in the remaining integral we can change $\vec
K$ to $-\vec K$ and using the fact that $C'_{\mu\nu\lambda\rho}(-k,q)
= - C'_{\mu\nu\lambda\rho}(k,q)$, we obtain
\begin{eqnarray}
\Gamma'^{(B1d)}_{\lambda\rho}(p) & = &
\frac{e^2}{2} \lim_{\vec Q\rightarrow 0}
\int\frac{d^3K}{(2\pi)^3 2K}f_\gamma(K)
\left(\frac{1}{k\cdot p}\right)\nonumber\\*
& & \times \Bigg\{
\gamma^\nu(\rlap/{p} + m)\gamma^\mu\left[
\frac{1}{2\vec K\cdot \vec Q - Q^2} - 
\frac{1}{2\vec K\cdot \vec Q + Q^2}\right]
[C'_{\mu\nu\lambda\rho}(k,q) -
C'_{\nu\mu\lambda\rho}(k,q)] 
\nonumber\\*
& & +
\gamma^\nu(-\rlap/{k})\gamma^\mu\left[
\frac{1}{2\vec K\cdot \vec Q - Q^2} + 
\frac{1}{2\vec K\cdot \vec Q + Q^2}\right]
[C'_{\mu\nu\lambda\rho}(k,q) + C'_{\nu\mu\lambda\rho}(k,q)]
\Bigg\} \nonumber\\[12pt]
& = & - 
\frac{e^2}{2} \lim_{\vec Q\rightarrow 0}
\int\frac{d^3K}{(2\pi)^3 2K}f_\gamma(K)
\frac{1}{k\cdot p} \left\{ {\gamma^\nu \rlap/k \gamma^\mu \over 
\vec K\cdot \vec Q}
\left[
C'_{\mu\nu\lambda\rho}(k,q) + C'_{\nu\mu\lambda\rho}(k,q)
\right]
+ O(Q)\right\} \,. \nonumber\\*
\end{eqnarray}
We now use
\begin{eqnarray}
(2v^\lambda v^\rho - \eta^{\lambda\rho})\gamma^\nu\rlap/{k}\gamma^\mu
\Big[ C'_{\mu\nu\lambda\rho}(k,q) +
C'_{\nu\mu\lambda\rho}(k,q) \Big] 
= 8(k\cdot v)(\vec K\cdot \vec Q) \rlap/v \,.
\end{eqnarray}
Then, using Eq.\ (\ref{ugu}) and putting $\vec P=0$, we get
\begin{eqnarray}
\label{B1Dresult}
m'_{(B2d)}
= - \frac{4e^2}{m}\int\frac{d^3K}{(2\pi)^3 2K}f_\gamma(K)
= -\frac{e^2 T^2}{6m} \,.
\end{eqnarray}

Adding the results given in Eqs. (\ref{B1Aresult}),
(\ref{B1Bresult}), (\ref{B1Cresult}) and (\ref{B1Dresult}), we get the
total contribution from the B1 term presented in Eq.\
(\ref{B1final}).

\subsection{The A2 contribution}\label{app:tough:A2}
For this contribution, we start from Eq.\ (\ref{A2}). Using Eq.\
(\ref{q&k}), we can write it as
\begin{eqnarray}
\Gamma'^{(A2)}_{\lambda\rho} (p,p) &=& {e^2 \over 2} 
\lim_{\vec Q \to 0} 
\int {d^4k \over (2\pi)^3} \; \delta(k^2-m^2) \eta_F(k) 
\nonumber\\* && \times 
\Bigg[ {\Lambda_{\lambda\rho} (k,k-q) 
\over (2\vec K\cdot \vec Q - Q^2) (m^2 - k\cdot p)} -
{\Lambda_{\lambda\rho} (k+q,k) \over (2\vec K\cdot \vec Q + Q^2)(m^2 -
k\cdot p')}
\Bigg] \,.
\end{eqnarray}
Following the strategy stated below Eq.\ (\ref{q&k}), let us now write
\begin{eqnarray}
\Lambda_{\lambda\rho} (k,k-q) &=& \Lambda_{\lambda\rho} (k,k) +
\Lambda'_{\lambda\rho} (k,q) \,, \nonumber\\* 
\Lambda_{\lambda\rho} (k+q,k) &=& \Lambda_{\lambda\rho} (k,k) + 
\Lambda''_{\lambda\rho} (k,q) \,,
\end{eqnarray}
and, in the denominator, expand $m^2-k\cdot p'$ in powers of $\vec Q$:
\begin{eqnarray}
{1 \over m^2-k\cdot p'} 
&=& {1 \over m^2-k\cdot p} + {\vec K \cdot \vec Q \over (m^2-k\cdot
p)^2} + O(Q^2) \,.
\end{eqnarray}
Then we can decompose $\Gamma'^{(A2)}_{\lambda\rho} (p,p)$ into the
following terms, omitting higher powers of $Q$ which anyway will not
contribute:
\begin{eqnarray}
\Gamma'^{(A2a)}_{\lambda\rho} (p,p) &=& {e^2 \over 2} 
\lim_{\vec Q \to 0} 
\int {d^4k \over (2\pi)^3} \; \delta(k^2-m^2) \eta_F(k) 
\nonumber\\* && \times 
{\Lambda_{\lambda\rho} (k,k) \over m^2 - k\cdot p} 
\Bigg(
{1 \over 2\vec K\cdot \vec Q - Q^2} - 
{1 \over 2\vec K\cdot \vec Q + Q^2}
\Bigg) \,, \nonumber\\ 
\Gamma'^{(A2b)}_{\lambda\rho} (p,p) &=& -\, {e^2 \over 4} 
\int {d^4k \over (2\pi)^3} \; \delta(k^2-m^2) \eta_F(k) 
{\Lambda_{\lambda\rho} (k,k) \over (m^2 - k\cdot p)^2} \,, 
\nonumber\\
\Gamma'^{(A2c)}_{\lambda\rho} (p,p) &=& {e^2 \over 2} 
\lim_{\vec Q \to 0} 
\int {d^4k \over (2\pi)^3} \; \delta(k^2-m^2) \eta_F(k) 
\nonumber\\* && \times 
{1 \over m^2 - k\cdot p} 
\Bigg(
{\Lambda'_{\lambda\rho} (k,q) \over 2\vec K\cdot \vec Q - Q^2} - 
{\Lambda''_{\lambda\rho} (k,q) \over 2\vec K\cdot \vec Q + Q^2}
\Bigg) \,, 
\end{eqnarray}
We discuss these contributions one by one.

\subsubsection*{The A2a contribution}
Using Eq.\ (\ref{Lambdayy}) and the $\delta$-function appearing in the
integrand, we can write
\begin{eqnarray}
\Gamma'^{(A2a)}_{\lambda\rho} (p,p) &=& -2e^2 
\lim_{\vec Q \to 0} 
\int {d^4k \over (2\pi)^3} \; \delta(k^2-m^2) \eta_F(k) 
\nonumber\\* && \times 
{k_\lambda k_\rho (\rlap/k - 2m) \over m^2 - k\cdot p} 
\Bigg(
{1 \over 2\vec K\cdot \vec Q - Q^2} - 
{1 \over 2\vec K\cdot \vec Q + Q^2}
\Bigg) \,.
\end{eqnarray}
As argued before Eq.\ (\ref{m'B1a}), we can put $\vec Q=0$ in the
spinors, and use Eq.\ (\ref{ugu}). Performing the $k_0$-integration,
we obtain
\begin{eqnarray}
(2v^\lambda v^\rho - \eta^{\lambda\rho}) 
\Gamma^{(A2a)}_{\lambda\rho}(p) &=& 
\frac{2e^2}{m}\lim_{\vec Q \rightarrow 0}
\int {d^3K \over (2\pi)^3} \; F(\vec K)
\Bigg(
{1 \over 2\vec K\cdot \vec Q - Q^2} - 
{1 \over 2\vec K\cdot \vec Q + Q^2}
\Bigg) \,,
\end{eqnarray}
where the expression on the left is understood to equal the one on the
right only between the spinors, and 
\begin{eqnarray}
F(\vec K) = {2E_K^2 - m^2 \over 2E_K}
\left[ \left( 1 + {m^2 \over m^2 - k\cdot p} \right) f_e + 
\left( 1 + {m^2 \over m^2 + k\cdot p} \right) f_{\bar e} \right] \,,
\end{eqnarray}
with $k_0=E_K$. Since the integrand contains $\vec K\cdot \vec P$, and
we must set $\vec P=0$ only after taking the limit $Q\to0$, the
angular integrations cannot be performed using Eq.\ (\ref{angint}). So
we shift the integration variable to $\vec K\pm\frac12\vec Q$ in the
terms having $2\vec K\cdot\vec Q \mp Q^2$ in the denominator. This
gives
\begin{eqnarray}
(2v^\lambda v^\rho - \eta^{\lambda\rho})
\Gamma^{(A2a)}_{\lambda\rho}(p) 
& = & \frac{2e^2}{m}
\lim_{\vec Q\rightarrow 0}\int\frac{d^3K}{(2\pi)^3}
\frac{\vec Q\cdot\vec\nabla_K F}{2\vec K\cdot\vec Q} \,.
\end{eqnarray}
Clearly the magnitude of $\vec Q$ now cancels out.  The derivative
with respect to $\vec K$ can be taken easily, using
\begin{eqnarray}
\vec\nabla_K E_K &=& {\vec K \over E_K} \,, \nonumber\\*
\vec\nabla_K\left(\frac{1}{m^2 \pm k\cdot p}\right) &=& 
\frac{\mp 1}{(m^2 \pm k\cdot p)^2}
\left(E_P\frac{\vec K}{E_K} - \vec P\right) \,.
\end{eqnarray}
The term proportional to $\vec P$ from the last derivative does not
contribute because it multiplies a factor whose integrand is odd in
$\vec K$ at $\vec P=0$.  Putting $\vec P=0$ in the other terms, we
obtain the contribution to the gravitational mass:
\begin{eqnarray}
m'_{(A2a)} = 
\frac{2e^2}{m}\int\frac{d^3K}{(2\pi)^3 2E_K} 
&\times& \Bigg\{ 
{2E_K^2-m^2 \over 2E_K} \bigg( {E_K-2m \over E_K-m} \; {\partial f_e \over
\partial E_K} + {E_K+2m \over E_K+m} \; {\partial f_{\bar e} \over
\partial E_K} \bigg) \nonumber\\* 
&& + {2E_K^2-m^2 \over 2E_K} \bigg( {m \over (E_K-m)^2} f_e - 
{m \over (E_K+m)^2} f_{\bar e} \bigg) \nonumber\\*
&& + {2E_K^2+m^2 \over 2E_K^2} \bigg( {E_K-2m \over E_K-m} f_e
+ {E_K+2m \over E_K+m} f_{\bar e} \bigg)
\Bigg\} \,.
\label{A2afinal}
\end{eqnarray}
%

\subsubsection*{The A2b contribution}
The integral in the (A2b) term is independent of $Q$.  So, in a
straight forward way, we obtain
\begin{eqnarray}
m'_{(A2b)} 
&=& {e^2 \over m^2} \int {d^3K \over (2\pi)^3 2E_K}
\; (2E_K^2 - m^2) \Bigg[ {E_K-2m \over (E_K-m)^2} f_e(E_K)
- {E_K+2m \over (E_K+m)^2} 
f_{\bar e}(E_K)  \Bigg] \,.
\label{A2bfinal}
\end{eqnarray}
%

\subsubsection*{The A2c contribution}
For the (A2c) contribution, first we use the expression for 
$\Lambda_{\lambda\rho}$ from Eq.\ (\ref{Lambda}) to find
\begin{eqnarray}
\Lambda'_{\lambda\rho} (k,q) &=& 
\eta_{\lambda\rho} (q^2-2k\cdot q) (\rlap/k-2m) + 
(k_\lambda q_\rho + k_\rho q_\lambda) (\rlap/k - 4m) 
+ k_\lambda \rlap/k \gamma_\rho \rlap/q 
+ k_\rho \rlap/k \gamma_\lambda \rlap/q \,,
\nonumber\\
\Lambda''_{\lambda\rho} (k,q) &=& 
\eta_{\lambda\rho} (q^2+2k\cdot q) (\rlap/k-2m)  
- (k_\lambda q_\rho + k_\rho q_\lambda) (\rlap/k - 4m) 
- k_\lambda \rlap/q \gamma_\rho \rlap/k 
- k_\rho \rlap/q \gamma_\lambda \rlap/k \,,
\end{eqnarray}
dropping irrelevant $O(q^2)$-terms and using $k^2=m^2$.  In the
$\eta_{\lambda\rho}$ terms, the integrand becomes independent of
$q$. Thus, these terms give a regular contribution. Let us denote it
by (A2r):
\begin{eqnarray}
m'_{(A2r)} &=& {2e^2 \over m} 
\int {d^3K \over (2\pi)^3 2E_K} \;
\Bigg[ {E_K-2m \over E_K-m} f_e(E_K) + 
{E_K+2m \over E_K+m} f_{\bar e}(E_K) \Bigg] \,.
\label{A2rfinal}
\end{eqnarray}

The terms which appear next will be called (A2s). For these, we use
the fact that
\begin{eqnarray}
(2v^\lambda v^\rho - \eta^{\lambda\rho}) (k_\lambda q_\rho + k_\rho
q_\lambda) = -2 k\cdot q = 2 \vec K \cdot \vec Q \,,
\end{eqnarray}
using $q\cdot v =q_0=0$. The $Q\to0$ limit can then be taken easily,
and we obtain
\begin{eqnarray}
m'_{(A2s)} 
&=& - {e^2 \over m} 
\int {d^3K \over (2\pi)^3 2E_K} \;
\Bigg[ {E_K-4m \over E_K-m} f_e(E_K) + 
{E_K+4m \over E_K+m} f_{\bar e}(E_K) \Bigg] \,.
\label{A2sfinal}
\end{eqnarray}

Finally, we come to the terms with three $\gamma$-matrices, which we
denote by (A2t). For these, first we note that
\begin{eqnarray}
(2v^\lambda v^\rho - \eta^{\lambda\rho}) 
(k_\lambda \rlap/k \gamma_\rho \rlap/q +
k_\rho \rlap/k \gamma_\lambda \rlap/q) = 
4 k\cdot v \rlap/k \rlap/v \rlap/q - 2m^2 \rlap/q \,,
\end{eqnarray}
and a similar expression with the other term. Since the $\rlap/q$ term
vanishes between the spinors, we can write
\begin{eqnarray}
m'_{(A2t)} &=& 2e^2 \lim_{P\to 0} \lim_{Q\to 0} \int {d^4k\over (2\pi)^3} 
\; \delta(k^2-m^2) \eta_F(k) \nonumber\\* && \times
{k_0 \over m^2 - k \cdot p} \; {1 \over 2 \vec K \cdot \vec Q} \; 
\overline u(p')
\Big( \rlap/k \rlap/v \rlap/q + \rlap/q \rlap/v \rlap/k \Big) u(p) \,,
\label{A2t}
\end{eqnarray}
omitting the $Q^2$ terms in the denominator since they will not
contribute for $Q\to0$.  Using the identity
\begin{eqnarray}
\gamma_\kappa \gamma_\mu \gamma_\nu 
= \eta_{\kappa\mu} \gamma_\nu + \eta_{\mu\nu}
\gamma_\kappa - \eta_{\kappa\nu} \gamma_\mu - i
\varepsilon_{\kappa\mu\nu\alpha} \gamma^\alpha \gamma_5 \,,
\end{eqnarray}
we obtain
\begin{eqnarray}
\rlap/k \rlap/v \rlap/q + \rlap/q \rlap/v \rlap/k 
= 2 \vec K\cdot \vec Q \rlap/v 
\end{eqnarray}
between the spinors, since $q\cdot v=0$ and $\rlap/q$ terms vanish.
Putting this back into Eq.\ (\ref{A2t}) and using Eqs.\ (\ref{ugu})
and (\ref{unorm}), we obtain
\begin{eqnarray}
m'_{(A2t)} = - {2e^2 \over m} 
\int {d^3K \over (2\pi)^3 2E_K} \;
\Bigg[ {E_K \over E_K-m} f_e(E_K) + 
{E_K \over E_K+m} f_{\bar e}(E_K) \Bigg] \,.
\label{A2tfinal}
\end{eqnarray}
The sum of Eqs.\ (\ref{A2afinal}), (\ref{A2bfinal}), (\ref{A2rfinal}),
(\ref{A2sfinal}) and (\ref{A2tfinal}) gives the total contribution of
the A2 term, given in Eq.\ (\ref{A2final}) in the text.

\subsection{The X contribution}\label{app:tough:X}
The part of the integral from Eq.\ (\ref{GamX.5}) that we consider
here is given by
\begin{eqnarray}
\label{Aq}
I^{(f)}(Q) 
& = & \int {d^3K\over (2\pi)^3} F(E_K) 
\left( {1\over 2 \vec K\cdot \vec Q - Q^2} -
{1\over 2 \vec K\cdot \vec Q + Q^2} \right) \,,
\end{eqnarray}
where
\begin{eqnarray} 
F(E) \equiv \Big(f_f(E) - f_{\bar f}(E) \Big)(2E^2 - m^2) \,.
\end{eqnarray}
Shifting the variables, the integral can be written as
\begin{eqnarray}
I^{(f)}(Q) & = &
\int {d^3K\over (2\pi)^3} 
{F(E_{\vec K + \frac{1}{2}\vec Q}) - 
F(E_{\vec K - \frac{1}{2}\vec Q}) \over 2\vec K\cdot\vec Q} \,. 
\end{eqnarray}
We have to expand the numerator to $O(Q^3)$ in order to obtain the
integral to $O(Q^2)$. Writing $\partial_i$ to denote a partial
derivative with respect to $K^i$,
\begin{eqnarray}
F(E_{\vec K \pm \frac{1}{2}\vec Q}) = F(E) \pm \frac{1}{2}Q^i\partial_i F
+ \frac{1}{2}\left(\frac{1}{4}Q^i Q^j\right)\partial_i\partial_j F \pm
\frac{1}{3!}\left(\frac{1}{8}Q^i Q^j Q^l\right)
\partial_i\partial_j\partial_l F \,.
\end{eqnarray}
The derivatives we need to use are:
\begin{eqnarray}
\partial_i F & = & K^i \left( {1\over E} {\partial \over \partial E}
\right) F \,, \nonumber\\
\partial_i\partial_j\partial_l F & = & (\delta^{ij}K^l + \delta^{il}K^j + 
\delta^{jl}K^i) \left( {1\over E} {\partial \over \partial E}
\right)^2 F +
K^i K^j K^l \left( {1\over E} {\partial \over \partial E} \right)^3 F \,.
\end{eqnarray}
Using
\begin{eqnarray}
K^i K^j \rightarrow \frac{1}{3}K^2\delta^{ij}
\end{eqnarray}
within the integrand, we have
\begin{eqnarray}
I^{(f)} (Q) = \int {d^3K \over (2\pi)^3}
\Bigg[ \frac12 \left( {1\over E} {\partial \over \partial E} \right) F + 
\frac{1}{3!}\frac{Q^2}{8}\left\{
3 \left( {1\over E} {\partial \over \partial E} \right)^2 F +  
\frac13 K^2 \left( {1\over E} {\partial \over \partial E} \right)^3 F 
\right\} \Bigg] \,.
\end{eqnarray}
Therefore, the quantity that we must substitute in Eq.\ (\ref{XQ-X0})
is
\begin{eqnarray}
I^{(f)}(Q) - I^{(f)}(Q\to 0) & = & 
\frac{1}{3!}\frac{Q^2}{8}\int {d^3K\over (2\pi)^3}
\left\{
3 \left( {1\over E} {\partial \over \partial E} \right)^2 F +  
\frac{K^2}{3} \left( {1\over E} {\partial \over \partial E} \right)^3 F 
\right\} \,.
\label{QX}
\end{eqnarray}
We now use the identity
\begin{eqnarray}
\int_0^\infty dK \; K^n \left( {1\over E} {\partial \over \partial E}
\right)^\nu F = 
- (n-1) \int_0^\infty dK \; K^{n-2} \left( {1\over E} {\partial \over
\partial E} \right)^{\nu-1} F \,,
\end{eqnarray}
which holds for $n\geq2$, so that the surface term vanishes. It is
obtained by using Eq.\ (\ref{EdE=KdK}) and performing a partial
integration.  Using it repeatedly, we can rewrite Eq.\ (\ref{QX}) as
\begin{eqnarray}
I^{(f)}(Q) - I^{(f)}(Q\to 0) & = & 
- \frac{Q^2}{48\pi^2}\int_0^\infty dK \;
\left( {1\over E} {\partial \over \partial E} \right) F \,.
\end{eqnarray}
Putting this back into Eq.\ (\ref{GamX.5}), we obtain the total X 
contribution given in Eq.\ (\ref{Xfinal}).

\end{document}